\documentclass[preprint]{aastex61}

\newcommand\aastex{AAS\TeX}

\usepackage{mathtools}

\submitjournal{ApJ}

\shorttitle{\aastex\ IR-radio correlation}
\shortauthors{Shi et al.}

\begin{document}

\title{The dependence of the IR-radio correlation on the metallicity}

\correspondingauthor{Yong Shi}
\email{yshipku@gmail.com}

\author{Jianjie Qiu}
\affil{School of Astronomy and Space Science, Nanjing University, Nanjing 210093, China.}
\affil{Key Laboratory of Modern Astronomy and Astrophysics (Nanjing University), Ministry of Education, Nanjing 210093, China.}

\author[0000-0002-8614-6275]{Yong Shi}
\affil{School of Astronomy and Space Science, Nanjing University, Nanjing 210093, China.}
\affil{Key Laboratory of Modern Astronomy and Astrophysics (Nanjing University), Ministry of Education, Nanjing 210093, China.}

\author{Junzhi Wang}
\affil{Shanghai Astronomical Observatory, Chinese Academy of Sciences, 80 Nandan Road, Shanghai 200030, China}

\author{Zhi-Yu Zhang}
\affil{Institute for Astronomy, University of Edinburgh, Royal Observatory, Blackford Hill, Edinburgh EH9 3HJ, UK}

\author{Luwenjia Zhou}
\affil{School of Astronomy and Space Science, Nanjing University, Nanjing 210093, China.}
\affil{Key Laboratory of Modern Astronomy and Astrophysics (Nanjing University), Ministry of Education, Nanjing 210093, China.}




\begin{abstract}

  We  have compiled a  sample of 26  metal-poor galaxies with  12 +
  log(O/H)  $<$ 8.1  with both  infrared continuum  and 1.4  GHz radio
  continuum data.  By comparing to  galaxies at higher metallicity, we
  have investigated the dependence on  the metallicity of the IR-radio
  relationship  at 24  $\mu$m, 70  $\mu$m, 100  $\mu$m and  160 $\mu$m
  bands as  well as the integrated  FIR luminosity.  It is  found that
  metal-poor  galaxies  have  on   average  lower  $q_{\rm  IR}$  than
  metal-rich ones with  larger offsets at longer  IR wavelengths, from
  -0.06 dex in $q_{\rm 24{\mu}m}$  to -0.6 dex in $q_{\rm 160{\mu}m}$.
  The q$_{\rm IR}$ of all galaxies as  a whole at 160 $\mu$m show
  positive trends with  the metallicity and IR-to-FUV  ratio, and 
  negative  trends  with  the  IR  color,  while  those  at  lower  IR
  wavelengths show weaker correlations.   We proposed a mechanism that
  invokes combined effects of  low obscured-SFR/total-SFR fraction and
  warm dust temperature at low metallicity to interpret the above behavior
  of $q_{\rm IR}$,  with the former reducing the IR
  radiation and the latter further  reducing the IR emission at longer
  IR  wavelength.  Other  mechanisms  that are  related  to the  radio
  emission  including   the  enhanced  magnetic  field   strength  and
  increased  thermal radio  contribution are  unable to  reconcile the
  IR-wavelength-dependent   differences   of  $q_{\rm   IR}$   between
  metal-poor and  metal-rich galaxies.  In contrast  to $q_{\rm IR}$,
  the mean total-SFR/radio ratio of metal-poor galaxies is the same as
  the metal-rich one,  indicating the 1.4 GHz radio  emission is still
  an effective tracer of SFRs at low metallicity.

\end{abstract}

\keywords{infrared/radio correlation, galaxy: mental poor, dwarf galaxy, galaxy: star formation}


\section{Introduction}
\label{sect:introduction}


A  tight linear  correlation between  radio emission  (at the
rest frequency  of 1.4  GHz) and  infrared (IR)  emission was
first  established  in  1980s   for  spiral  galaxies  \citep{Helou85,
  deJong85}.  The  infrared emission is  the re-radiation of  the dust
heated by  UV radiation of  massive stars.  The radio emission  is the
synchrotron  radiation  of  cosmic  ray electrons  as  accelerated  by
shocks in  Type II  supernova (SN)  remnants \citep{Helou93}.
At high radio  frequency, the thermal radio emission  from HII regions
could also be  an important component.  As both IR  and radio emission
originate  from star  formation, the  relation is  widely utilized  to
study star formation activities \citep{Condon02, Murphy06a, Murphy06b}
as well  as an effective  means to distinguish between  star formation
galaxies and AGN \citep{Yun01}.

The IR-radio correlation spans five  orders of magnitude in luminosity
and holds for both late-type field star-forming galaxies \citep{Yun01}
and low-mass Magellanic-type peculiar galaxies \citep{Jurusik14}.  Not
only valid in the local universe, it also holds at high redshift up to
z  $\sim$ 1  - 3  \citep{Appleton04,Sargent10,Ivison10a}, although  at
high $z$ in addition to  the Synchrotron radiation, the  energy loss
due to the inverse Compton scatter off the cosmic microwave background
may be important \citep{Murphy09b}.   Even with the stacking technique
(See  \citealt{Ivison10a}  for details),  many  works  still found  no
redshift   evolution   of   the    correlation   out   to   $z$ $\sim$ 3
\citep{Ivison10a, Ivison10b, Sargent10, Mao11}.

However,   outliers  to   the  relation   do  exist,   including  both
radio-excess  and radio-deficient  (or FIR-excess)  ones.  The  global
FIR/radio ratio of cluster galaxies is  found to be lower than that of
field galaxies  \citep{Miller01, Murphy09a}.  It is  proposed that the
interaction  between  ISM  and   ICM  produces  shocklets  to  further
accelerate CR electrons  in the galaxy, enhancing  the radio emission.
Such  radio-excess  sources are  also  found  in the  massive  cluster
MS0451.6-0305  at   z  $\sim$  0.54   \citep{Randriamampandry15}.   To
investigate   the  effect   of  tidal   shocks  created   by  galactic
interactions  and mergers,  \citet{Donevski15}  studied  the trend  of
FIR/radio ratio  with different merging stages  for 43 infrared-bright
star-forming interacting galaxies, and found a notable radio excess at
some merger  stages. They proposed that  the radio excess is  not only
caused by the  non-thermal radio emission from the  gas bridge between
interacting galaxies,  but also  by the emission  of the  CR electrons
accelerated by shocks within  individual interacting galaxies.  At
high-$z$,  \citet{Smolcic15}  found  radio  excess  in  sub-millimeter
galaxies  at $z$  $\sim$  4 -  6,  which  is argued  to  be caused  by
selection effects as  these sources are at early  stages of evolution.
On  the other  hand,  the radio-deficient  galaxies  are likely  those
star-bursts at  very young stages where  SN has not exploded  yet while
the dust emission could be present \citep{Roussel03}.

In the past  decade, due to the improved spatial  resolution of the IR
and radio  observations, studies  of the  IR-radio relations  are also
carried out for  spatially resolved nearby spiral  galaxies.  A series
of   works   for   a   few   dozens   of   galaxies   were   done   by
\citet[etc.]{Murphy06a,Murphy06b,Murphy08}  based on  the observations
with {\it  Spitzer} Infrared Telescope and  Westerbork Synthesis Radio
Telescope  (WSRT).    By  using  the  image-smearing   technique  (See
\citealt{Murphy06a} for details), they studied the cosmic-ray electron
diffusion  at  sub-kpc scale,  and  showed  that the  FIR/radio  ratio
decreases with  both the  declining surface brightness  and increasing
radius.

Most  of  the  previous  studies,  however,  focused  on  star-forming
galaxies   around  solar   metal   abundance,  with   few  works   for
low-metallicity dwarf galaxies. The early investigations based on IRAS
60 $\mu$m  and VLA 1.4  GHz detection of  about 15 blue  compact dwarf
galaxies around 12 + log(O/H) = 8.0 indicate that the ratio of the two
does not  show systematic offsets from  the spirals \citep{Hopkins02},
and the  derived SFRs from the  two wavelengths agree with  each other
over five  order of  magnitudes. Investigations of  a small  sample of
local group dwarfs did not find the offset in the $f_{60{\mu}m}/f_{\rm
  2.64GHz}$ ratio from spiral galaxies either \citep{Chyzy11}. Another
work about dwarf galaxies with {\it Spitzer} 24 $\mu$m and VLA 1.4 GHz
detection  shows the  deviation of  q$_{24}$ at  low metallicity  from
spirals \citep{Wu08}.  They  found that above 12 + log(O/H)  = 8.0 the
q$_{24}$ is almost  constant while below 12 + log(O/H)  = 8.0 the five
out of six detection follows  the decreasing trend with the decreasing
metallicity, except for an outlier  (SBS 0335-052E) that is even above
the bulk of those spirals.

Metal-poor dwarf galaxies are far  more numerous than massive galaxies
and serves  as the  building blocks of  massive galaxies  at high-$z$.
Extremely metal-poor  galaxies (12 +  log(O/H) $<$ 7.6) also  offer an
opportunity to study the galaxy  evolution at the quasi-pristine metal
environment \citep{Shi14, Shi15, Shi16}.  The IR-radio relation offers
a  powerful  way  to  understanding a  series  of  physical  processes
including star formation, dust heating, cosmic ray, magnetic field etc
that   happen   in  the   interstellar   medium   (ISM)  of   galaxies
\citep[e.g.]{Schleicher16}.   We  would  like to  study  the  IR-radio
relation  of metal-poor  galaxies  and gain  insights  into the  above
physical processes in these galaxies. We will compile as many as dwarf
galaxies  with  12 +  log(O/H)  $\leq$  8.1  in  the archive  of  {\it
  Herschel} Space Observatory  as well as previous  IR missions, along
with the  radio data  from the  literature.  The goal  is to  not only
enlarge the  dwarf sample  to study the  IR/radio relationship  at low
metallicity but also to investigate the IR/radio ratio at different IR
wavelengths.  In  \S~\ref{sect:sample}  we   present  the  sample  and
data. The  results are shown in  \S~\ref{sect:result}. The discussions
are       presented       in       \S~\ref{sect:discussion}.        In
\S~\ref{sect:conclusion} we present the conclusions.

\section{Sample Selection and Data Reduction}
\label{sect:sample}

\subsection{Sample Selection}

\begin{figure}
\epsscale{1.0}    
\plotone{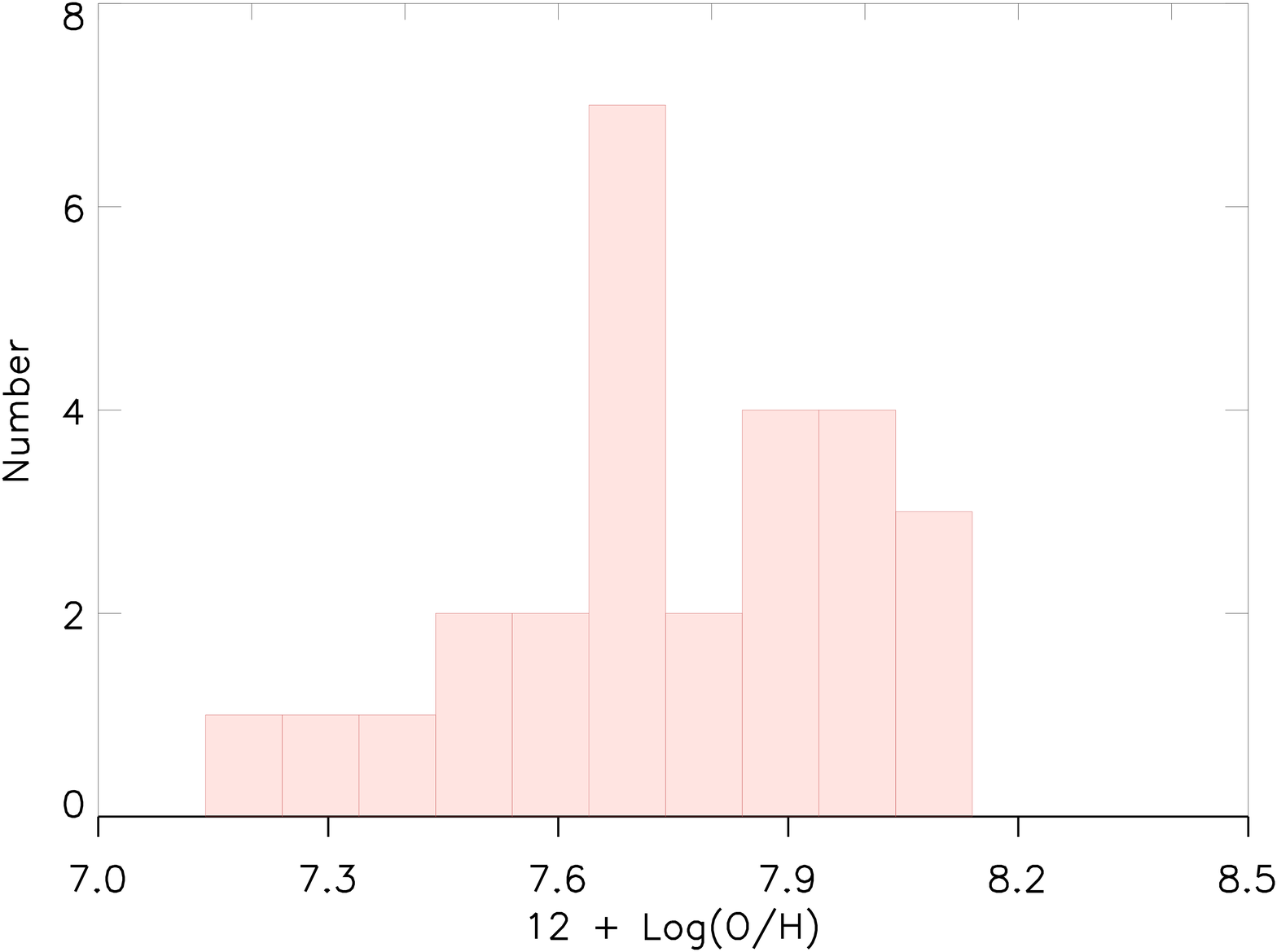}
   \caption{The distribution of the oxygen abundance 12 + log(O/H) of our compiled metal-poor galaxies.} 
   \label{Fhistogram}
   \end{figure}

In order to construct a sample  involving as many metal-poor dwarfs as
possible, we went  through all programs in the  {\it Herschel} science
archive  of  nearby  galaxies  ($z<$0.1) and  compiled  a  catalog  of
galaxies that have broad-band images as observed by {\it Herschel}. We
then searched for the measurements  of the oxygen nebular abundance in
the literature and defined our metal-poor sample as galaxies with 12 +
log(O/H) $\leq$ 8.1.  The  metallicity measurements of this metal-poor
sample  are mainly  based on  the  direct method,  while a  comparison
sample of metal-rich galaxies mainly uses various strong line methods.
The former  has a high precision  (0.1 dex) in contrast  to the latter
with     large     systematic    uncertainties     ($\sim$0.5     dex)
\citep{Moustakas10}.  Since our study  focuses on the metal-poor ones,
we expect the results are  not affected significantly by the abundance
measurement  error.   To further  increase  the  number of  metal-poor
galaxies,   we   included   dwarf  objects   from   \citet{Wu08}   and
\citet{Klein91}.  The radio 1.4 GHz  continuum data were compiled from
the  literature through  NED and  VLA  archive.  The  final sample  as
listed in Table~\ref{table_dwarfsample} contains 26 galaxies with 12 +
log(O/H) $\leq$ 8.1. Figure \ref{Fhistogram} shows the distribution of
the oxygen abundance of these galaxies.

For the high-metallicity comparison sample, we included galaxies above
12  + log(O/H)  = 8.1  from  Spitzer Infrared  Nearby Galaxies  Survey
(SINGS)  \citep{Dale07,  Moustakas10},  the  Key  Insights  on  Nearby
Galaxies:  A  Far-Infrared  Survey  with  Herschel  sample  (KINGFISH)
\citep{Dale12}    and   Dwarf    Galaxies   Survey    (DGS)   programs
\citep{Madden13}.   We  removed  AGN identified  through  the  optical
emission line  diagnostics and radio-loud AGN  by \citet{Moustakas10},
\citet{Mendoza15}  and \citet{Best12}.   The presence  of faint  radio
emission  from  the central  black-holes  is  still possible  in  this
comparison   sample,  but   because   our   sample  is   well-resolved
star-forming galaxies so that such emission should be a small fraction
of  the total  radio emission  and thus  the $q$  value should  not be
affected.     The   final    comparison    sample    is   listed    in
Table~\ref{table_comparionsinfrared}.

To  reduce the systematic  uncertainties in the  IR photometry
measurements for our metal-poor  sources that are generally faint
in  the IR,  we carried  out the  aperture photometry  of the  IR
fluxes  at 24,  70,  100, and  160 $\mu$m.   The  IR images  were
retrieved from the data archives of two telescopes including {\it
  Spitzer} Space Telescope and  {\it Herschel} Space Observatory.
These  observations   were  mainly   done  in  the   programs  of
\citet{Dale07,   Dale09,    Dale12}   and   \citet{Remy-Ruyer13}.
Aperture loss was  corrected based on the  point spread functions
of  each telescope  at the  corresponding wavelength.   The final
derived IR  fluxes listed in  Table~\ref{table_dwarfinfrared} are
consistent with the literature values within 20\%, 25\%, 29\% and
41\% at 24,  70, 100, and 160 $\mu$m, respectively.   As Mrk 1499
has no {\it  Spitzer} or {\it Herschel} data so  that the IRAS 60
and 100 $\mu$m photometry were used to interpolate the 70 and 100
$\mu$m fluxes.   For metal-rich sources, the  IR photometry were
   collected  from the  literature.   The GALEX  far-UV  data of  both
   samples were retrieved  from the NED. All photometry  are listed in
   Table~\ref{table_dwarfsample},  Table~\ref{table_dwarfinfrared} and
   Table~\ref{table_comparionsinfrared}.

\subsection{Measurements of $q$ parameter, SFRs }

Following \citet{Helou85}, the IR-to-radio ratio is defined as $q_{\rm
  IR}$ = log$(\frac{S_{\rm IR }}{S_{1.4\rm GHz}})$, where $S_{\rm IR}$
is the monochromatic IR flux at a given IR wavelength or the total FIR
flux, and  $S_{1.4 \rm  GHz}$ is  the 1.4  GHz radio  continuum
emission. Both fluxes are in the unit of Jy.  The total FIR luminosity
is measured by $L_{\rm{FIR}}= 4.63 \times 10^{-15} \times (8.3L_{24} +
2.7L_{70} + L_{160})$, where $L_{\rm{FIR}}$  was in unit of $L_{\sun}$
with     $L_{24}$,     $L_{70}$,     and     $L_{160}$     in     W/Hz
\citep{Symeonidis08}. The FIR luminosity was  then divided by
  the  median  frequency  to  have   the  FIR  flux  $S_{\rm  FIR}$  =
  $\frac{L_{\rm FIR}}{4\pi \times D^2 \times \nu_{\rm 85 \mu m}}$ in Jy.

The SFR is derived by using the equation
\begin{equation}
{\rm SFR} =  0.68\, {\times}\, 10^{-28}L_\nu ({\rm FUV}) + 2.14\, {\times}\, 10^{-42}L({24\,{\mu}m)}
\end{equation}
, where   the  $L_\nu ({FUV})$   is  the   FUV  band   luminosity  in
erg s$^{-1}$ Hz$^{-1}$, the $L({24\, {\mu}m})$ is the 24 $\mu m$ band
luminosity in erg  s$^{-1}$, and the  unit of  SFR is $M_{\odot}$ yr$^{-1}$ \citep{Leroy08}.
The calibration of SFRs have systematic uncertainties up to about
  0.3   dex.   The   initial   mass  function   may   change  at   low
  metallicity. The low  opacity due to the low  metallicity could also
  result in stronger  UV radiation for massive stars so  that the SFRs
  of metal-poor galaxies may be overestimated. However, since our main
  focus is about the $q$ parameter  which is the ratio of two observed
  fluxes, the systematic uncertainty of the SFR measurement should not
  affect our main results.

\section{Result}
\label{sect:result}

\subsection{$q_{\rm IR}$ parameter as a function of the oxygen abundance} 

\begin{figure}
  \epsscale{0.60}
  \plotone{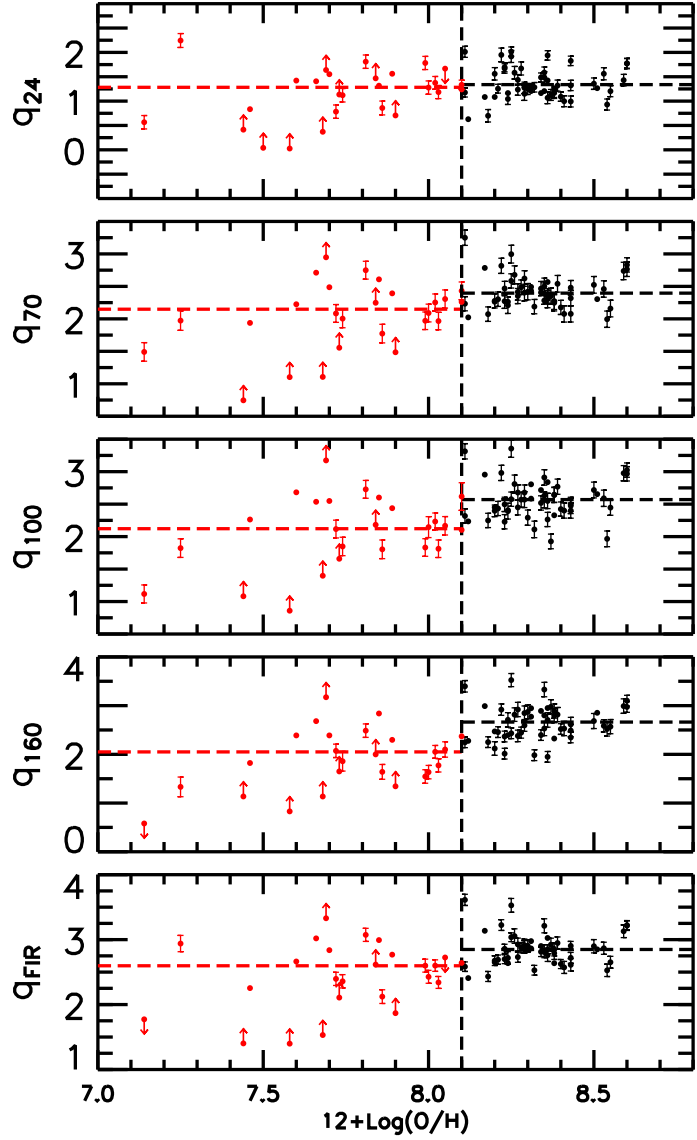}
  \caption{The q$_{\rm IR}$  as a function of the  oxygen abundance at
    24$\mu$m, 70$\mu$m,  100$\mu$m, 160$\mu$m, and FIR  (definition at
    Section  \ref{sect:sample}).  The  red  points  denote  metal-poor
    galaxies  and  black  points  are for  metal-rich  galaxies.   The
    vertical dashed line  represents the value of 12  + log(O/H) = 8.1,
    while two horizontal  dashed lines indicate the  average values of
    metal-poor and metal-rich galaxies, respectively. \label{F_qvsM}}
\end{figure}

Figure \ref{F_qvsM} shows  the trend of $q_{\rm IR}$ as  a function of
the gas-phase oxygen abundance,  including $q_{\rm 24{\mu}m}$, $q_{\rm
  70{\mu}m}$, $q_{\rm  100{\mu}m}$ and $q_{\rm 160{\mu}m}$.   The mean
$q_{\rm 24{\mu}m}$ of our metal-rich  (12 + log(O/H) $>$ 8.1) galaxies
is  1.34$\pm$0.05 as  listed  in  Table~\ref{table_linefit} where  the
uncertainty is  the error of the  mean. This average is  comparable to
the literature  value in \citet{Wu08} but  larger by 0.3 dex  than the
result of \citet{Appleton04}.  As shown in the first panel, metal-poor
galaxies with  12 + log(O/H)  $<$ 8.1  have an average  $q$, excluding
lower-limits, close to  the mean $q$ of our  metal-rich galaxies.  The
entire sample shows  no apparent trend between  $q_{\rm 24{\mu}m}$ and
the metallicity with a Kendalls' rank correlation coefficiency of only
0.16, inconsistent  with what is  found in \citet{Wu08} who  claimed a
rough  trend  of  decreasing  $q_{\rm  24{\mu}m}$  with  the  reducing
metallicity.   As  we  included  all sources  from  \citet{Wu08},  the
difference  is most  likely  caused  by the  small  sample studied  in
\citet{Wu08}.  Below  12 + log(O/H) $<$  7.6, there is some  hint that
the dispersion  increases but the  number of galaxies is  very limited
mainly because of few radio  continuum detection of such galaxies. The
$q_{\rm 24{\mu}m}$ of the extremely metal-poor galaxy SBS 0335-052E is
boosted due  to its prominent warm  dust emission as its  IR SED peaks
around 24  $\mu$m, while  another extremely  metal-poor galaxy  IZw 18
instead has a value below the mean.

In  the $q_{\rm  70{\mu}m}$ panel,  the mean  value of  our metal-rich
sample is  close to the mean  of the sample by  \citet{Yun01}, still a
bit  larger  than  the  value by  \citet{Appleton04}.   As  listed  in
Table~\ref{table_linefit},  the   mean  $q_{70}$  of   our  metal-poor
galaxies was lower  than the mean of metal-rich  ones by 0.26$\pm$0.09
dex.  Especially, SBS 0335-052E  that shows exceptionally high $q_{\rm
  24{\mu}m}$ now has  $q_{\rm 70{\mu}m}$ lower than the  mean value of
metal rich  galaxies (12 +  log(O/H) $>$ 8.1).  All  galaxies together
show  a weak  trend  of  $q_{\rm 70{\mu}m}$  that  decreases with  the
reducing metallicity, with a correlation coefficiency of 0.32.

The overall behavior of $q_{\rm 70{\mu}m}$  is also seen in the panels
of $q_{\rm 100{\mu}m}$  and $q_{\rm 160{\mu}m}$. Compared  to the case
of  $q_{\rm  70{\mu}m}$,  the  mean $q_{\rm  100{\mu}m}$  and  $q_{\rm
  160{\mu}m}$ of metal-poor galaxies are increasingly lower than those
of  metal-rich  ones,   by  0.44$\pm$0.12  and 0.61$\pm$0.13  dex,
  respectively.  At both wavelengths, there exists rough correlations
between $q$ and  the metallicity for all galaxies  together, with
  correlation coefficiencies  of 0.42 and  0.57 at 100 $\mu$m  and 160
  $\mu$m, respectively.  The last  panel of Figure \ref{F_qvsM} shows
that mean $q_{\rm FIR}$ of metal-poor  galaxies is also lower than the
mean value of metal-rich ones by 0.24$\pm$0.09 dex.

\subsection{$q_{\rm IR}$ parameter as a function of the IR luminosity and IR-to-FUV ratio}

The drop in the $q$ parameter  for low-metallicity galaxies could be a
result of the low dust content in  these galaxies so that only a small
fraction of  radiation from massive  stars is reprocessed by  dust and
re-emitted in the IR bands.

\begin{figure}
\epsscale{1.0}    
\plotone{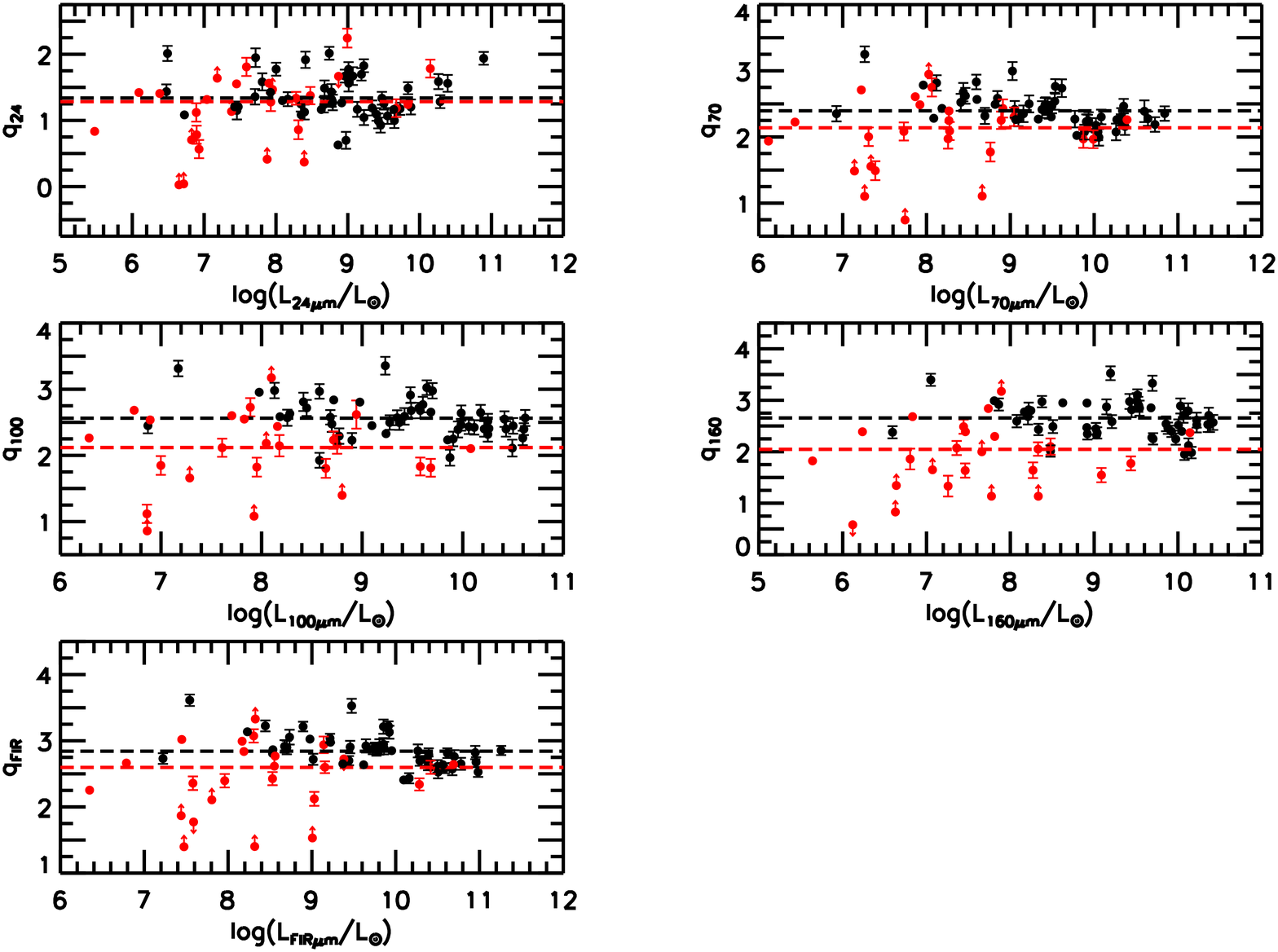}
\caption{  The  q$_{\rm IR}$  as a function  of the IR luminosity.  The red  points
  denote  metal-poor  galaxies and  black  points  are for  metal-rich
  galaxies.   The two  horizontal  dashed lines  indicate the  average
  values of metal-poor and metal-rich galaxies, respectively.  \label{F_qvsL}}
\end{figure}

We first show the  $q_{\rm IR}$ as a function of  the IR luminosity in
Figure~\ref{F_qvsL}.   At  24  $\mu$m, metal-poor  galaxies  are  less
luminous  than metal-rich  ones but  still  cover a  large range  from
around 10$^{5}$  to 10$^{10}$  $L_{\odot}$, while the  metal-rich ones
have a  luminosity range from  10$^{6}$ to 10$^{11}$  $L_{\odot}$.  At
three longer wavelengths, metal-poor galaxies also cover a large range
from 10$^{6}$ to 10$^{10}$ $L_{\odot}$.  Overall, there is no apparent
dependence of  $q_{\rm IR}$  on the IR  luminosity, indicating  the IR
luminosity should  not be the main  driver of lower $q$  of metal-poor
galaxies.  \citet{Bell03}  investigated the  dependence of $q$  of 249
galaxies on the IR luminosity  down to 10$^{8}$ $L_{\odot}$, and found
no  systematic change  in the  $q$, consistent  with our  results that
further extends the study to 10$^{5}$-10$^{6}$ $L_{\odot}$.

\begin{figure}
\epsscale{0.70}    
\plotone{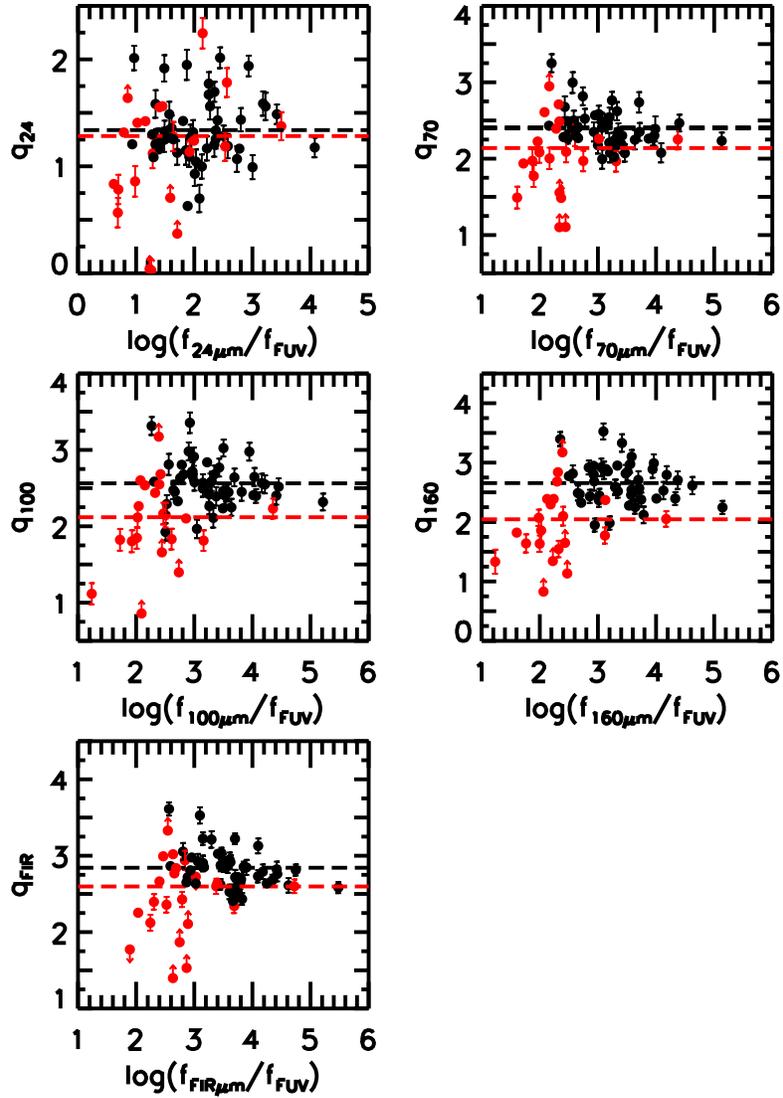}
\caption{The  q$_{\rm IR}$  as a function  of the
  IR-to-FUV  ratio at  corresponding  IR wavelength.   The red  points
  denote  metal-poor  galaxies and  black  points  are for  metal-rich
  galaxies.   The two  horizontal  dashed lines  indicate the  average
  values of metal-poor and metal-rich galaxies, respectively. \label{F_qvsLIR_FUV}}
\end{figure}

Figure~\ref{F_qvsLIR_FUV} shows the $q_{\rm IR}$  as a function of the
IR  to far-UV  flux ratio  at  different IR  wavelengths.  On  average
metal-poor  galaxies  show  lower   IR/far-UV  ratio  than  metal-rich
galaxies,  indicating lower  dust extinction  in metal-poor  galaxies.
There is  a weak trend  of decreasing $q_{\rm 24{\mu}m}$  with reduced
$f_{\rm  24{\mu}m}/f_{\rm   FUV}$  for  the  entire   sample,  with  a
correlation coefficiency  of 0.22.  Such a behavior  is also
  seen in $q_{\rm 100{\mu}m}$ vs.  $f_{\rm 100{\mu}m}/f_{\rm FUV}$ and
  $q_{\rm  160{\mu}m}$  vs.   $f_{\rm  160{\mu}m}/f_{\rm  FUV}$,  with
  correlation   coefficiencies  of   0.25  and   0.42,  respectively.
Although there  is no  trend between  $q_{\rm 70{\mu}m}$  vs.  $f_{\rm
  70{\mu}m}/f_{\rm FUV}$,  metal-poor galaxies  on average  have lower
$f_{\rm  70{\mu}m}/f_{\rm   FUV}$  than  metal-rich   galaxies.  These
dependencies on the IR/FUV ratio indicates that the lower $q_{\rm IR}$
of  metal-poor galaxies  are associated  with their  smaller IR/far-UV
ratio at the corresponding IR wavelength.

\subsection{$q_{\rm IR}$ parameter as a function of the IR color}

The  IR SED  is sensitive  to many  physical parameters  such as  dust
temperature, the heating  source and dust grain properties  etc. It is
already  known  that  the  dust  IR SED  of  metal-poor  galaxies  are
different   from    those   metal-rich    ones   \citep{Engelbracht08,
  Remy-Ruyer13,  Shi14,  Zhou16}.    Previous  investigations  of  the
dependence of  the $q_{\rm  IR}$ value  on the IR  color do  not reach
conclusive   results.  For   example,   no  trend   is  seen   between
$q_{24{\mu}m}$   vs.   f(60$\mu$m)/f(100$\mu$m)   in   the   work   of
\citet{Wu08},  while the  dependence has  been  found in  the work  of
\citet{Hummel88} for  the $q_{100{\mu}m}$  as a  function of  the dust
temperature,  and in  the work  of \citet{Roussel03}  for the  $q_{\rm  FIR}$
with f(60$\mu$m)/f(100$\mu$m).

\begin{figure}
\epsscale{0.70}    
\plotone{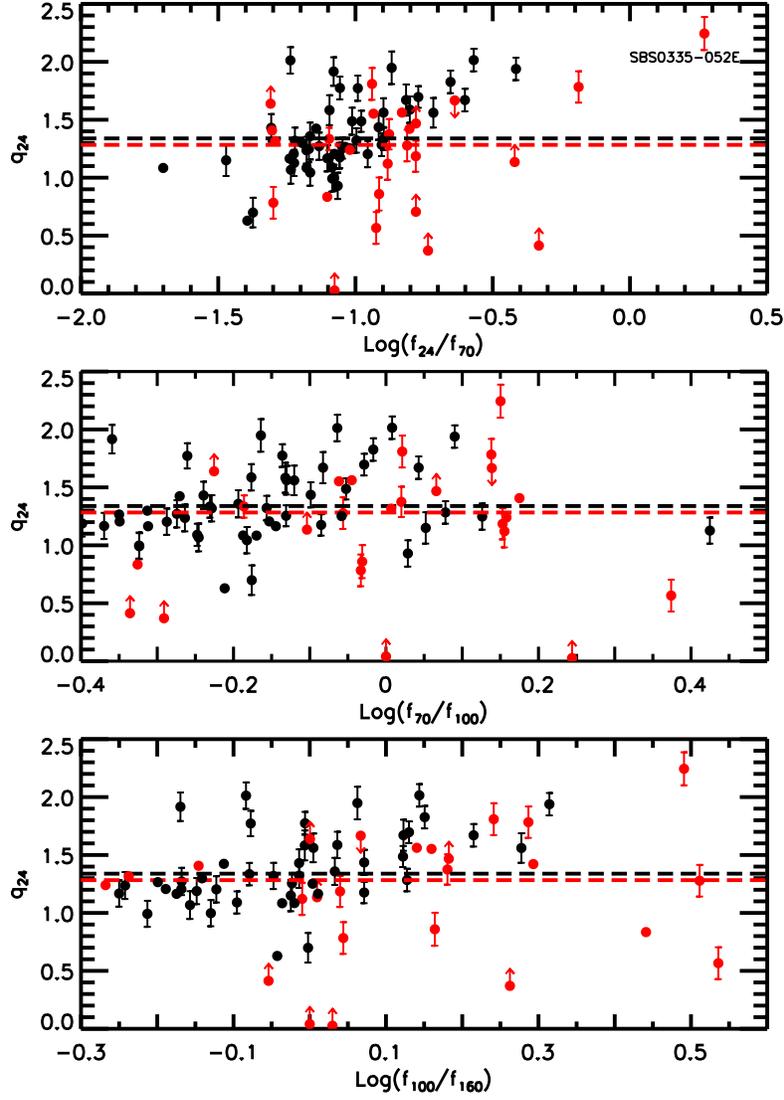}
\caption{ q$_{\rm 24{\mu}m}$ vs. the IR color of  $f_{24{\mu}m}/f_{70{\mu}m}$,
$f_{70{\mu}m}/f_{100{\mu}m}$ and $f_{100{\mu}m}/f_{160{\mu}m}$.  The  red  points  denote  metal-poor
    galaxies  and  black  points  are for  metal-rich  galaxies.  The two horizontal  dashed lines indicate the  average values of
    metal-poor and metal-rich galaxies, respectively.    \label{Fcolor24}}
\end{figure}

Figure~\ref{Fcolor24}  plots  $q_{24}$  vs.   IR  color  at  different
wavelengths           including           $f_{24{\mu}m}/f_{70{\mu}m}$,
$f_{70{\mu}m}/f_{100{\mu}m}$ and $f_{100{\mu}m}/f_{160{\mu}m}$. Except
for     SBS     0335-052E,      the     short     wavelength     color
$f_{24{\mu}m}/f_{70{\mu}m}$  of metal-poor  galaxies bears  an overall
similar  range to  that of  metal-rich  ones.  A weak  positive
  correlation is seen between $q_{24}$ and $f_{24{\mu}m}/f_{70{\mu}m}$
  for the  entire sample,  with a  correlation coefficiency  of 0.33.
The  second  panel  of  Figure~\ref{Fcolor24}  shows  that  metal-poor
galaxies  have on  average  warmer $f_{70{\mu}m}/f_{100{\mu}m}$  color
than metal rich ones but  no apparent correlation between $q_{24}$ and
$f_{70{\mu}m}/f_{100{\mu}m}$.  Actually the  two galaxies with highest
$f_{70{\mu}m}/f_{100{\mu}m}$ around  0.4 show  almost the  largest and
smallest $q_{24}$, respectively.  The above result confirms a previous
finding     of    no     relation    between     $q_{24{\mu}m}$    and
f(60$\mu$m)/f(100$\mu$m)  \citep{Wu08}.    The  last   panel  of
  Figure~\ref{Fcolor24}   illustrates   some    weak   dependence   of
  $q_{24{\mu}m}$  on the  far-IR color  $f_{100{\mu}m}/f_{160{\mu}m}$,
  with a correlation coefficiency of 0.26.

\begin{figure}
\epsscale{0.70}    
\plotone{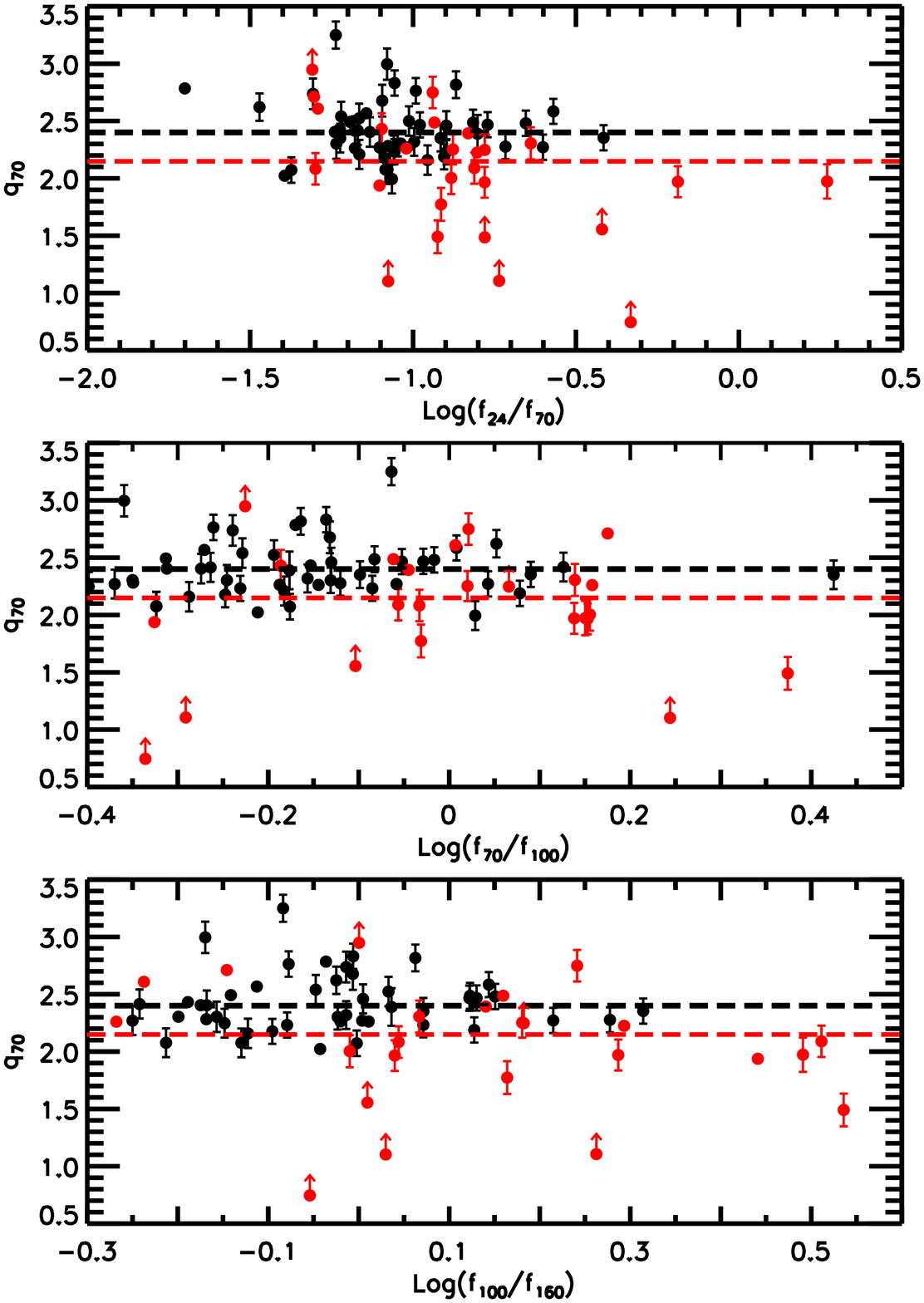}
\caption{ The same as Fig.~\ref{Fcolor24} but for  q$_{\rm 70{\mu}m}$.\label{Fcolor70}}
\end{figure}

Figure~\ref{Fcolor70} investigates the dependence of $q_{70{\mu}m}$ on
the  above three  IR color. A weak correlation  is seen  between
$q_{70{\mu}m}$   and   $f_{24{\mu}m}/f_{70{\mu}m}$,      with   a
  correlation  coefficiency of  -0.34.  Overall  the metal-poor  and
metal-rich      galaxies     share      the     same      range     in
$f_{24{\mu}m}/f_{70{\mu}m}$ as already mentioned in the above.  At the
same $f_{24{\mu}m}/f_{70{\mu}m}$,  $q_{70}$ is smaller  for metal-poor
galaxies  as  compared to  metal-rich  galaxies.   The two  metal-poor
galaxies  with highest  $f_{24{\mu}m}/f_{70{\mu}m}$ show  low $q_{70}$
but not as extreme as expected if there is a relation.  The dependence
of  $q_{70}$ on  the $f_{70{\mu}m}/f_{100{\mu}m}$  is very  weak, 
  with  a correlation  coefficiency  of -0.05.   At  the same  color,
metal-poor galaxies on average  occupy lower $q_{70{\mu}m}$ regimes as
compared  to  metal-rich ones.   Among  two  metal-poor galaxies  with
highest  $f_{70{\mu}m}/f_{100{\mu}m}$, only  one shows  lower $q_{70}$
than  the remaining  all galaxies.   The  last panel  shows no  strong
relation    between    $q_{70{\mu}m}$    and    the    far-IR    color
$f_{100{\mu}m}/f_{160{\mu}m}$, with a correlation coefficiency of
  about -0.2.

\begin{figure}
\epsscale{0.70}    
\plotone{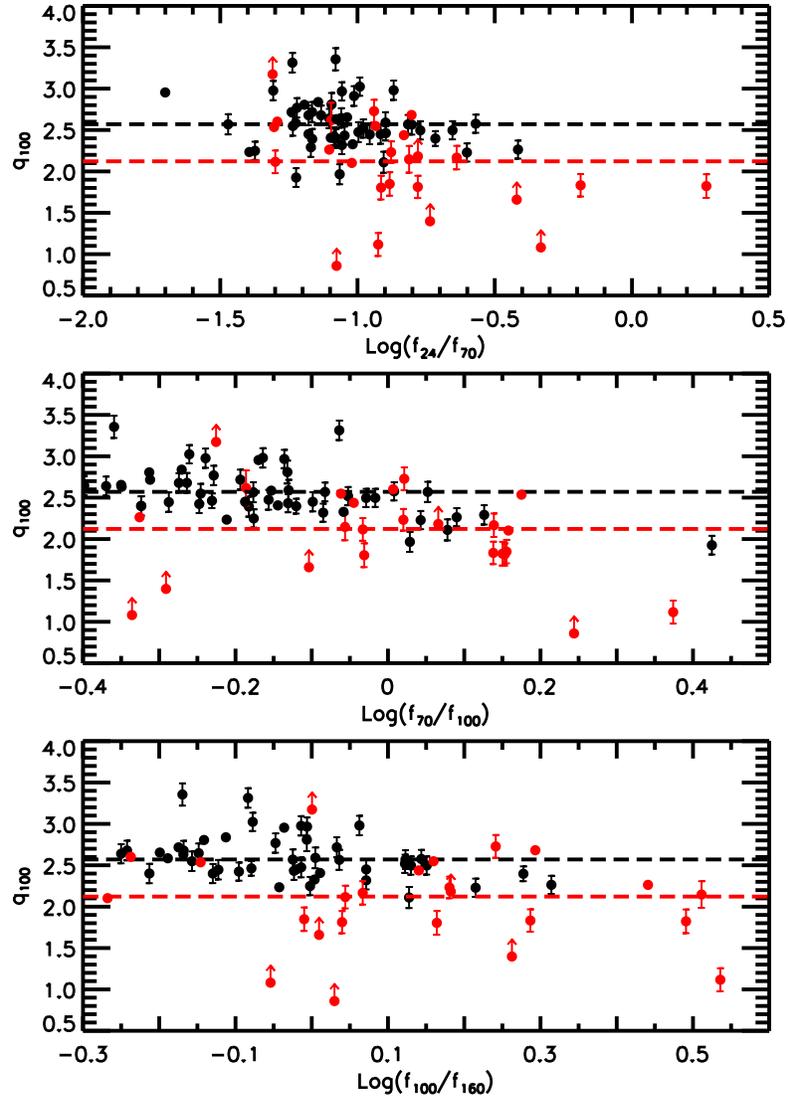}
\caption{The same as Fig.~\ref{Fcolor24} but for  q$_{\rm 100{\mu}m}$.\label{Fcolor100}}
\end{figure}

Figure~\ref{Fcolor100} shows the $q_{100{\mu}m}$  as a function of the
above three IR  color. The first panel indicates a  weak dependence of
$q_{100}$  on  the  $f_{24{\mu}m}/f_{70{\mu}m}$,  with  a  correlation
coefficiency of  -0.42.  At the  same color, metal-poor  galaxies show
somewhat  lower $q_{100{\mu}m}$  than  the metal  rich  ones. The  SBS
0335-052E with  highest $f_{24{\mu}m}/f_{70{\mu}m}$ does not  show the
lowest $q_{100}$.  In the second  panel some stronger trend is present
for the $q_{100{\mu}m}$ with  the $f_{70{\mu}m}/f_{100{\mu}m}$, with a
correlation coefficiency of -0.53.  The  two galaxies with the highest
$f_{70{\mu}m}/f_{100{\mu}m}$ do have lowest  $q_{100}$ among the whole
sample.  In  the  last  panel,  the dependence  of  $q_{100}$  on  the
$f_{100{\mu}m}/f_{160{\mu}m}$ is weak  with a correlation coefficiency
of -0.33.

\begin{figure}
\epsscale{0.70}    
\plotone{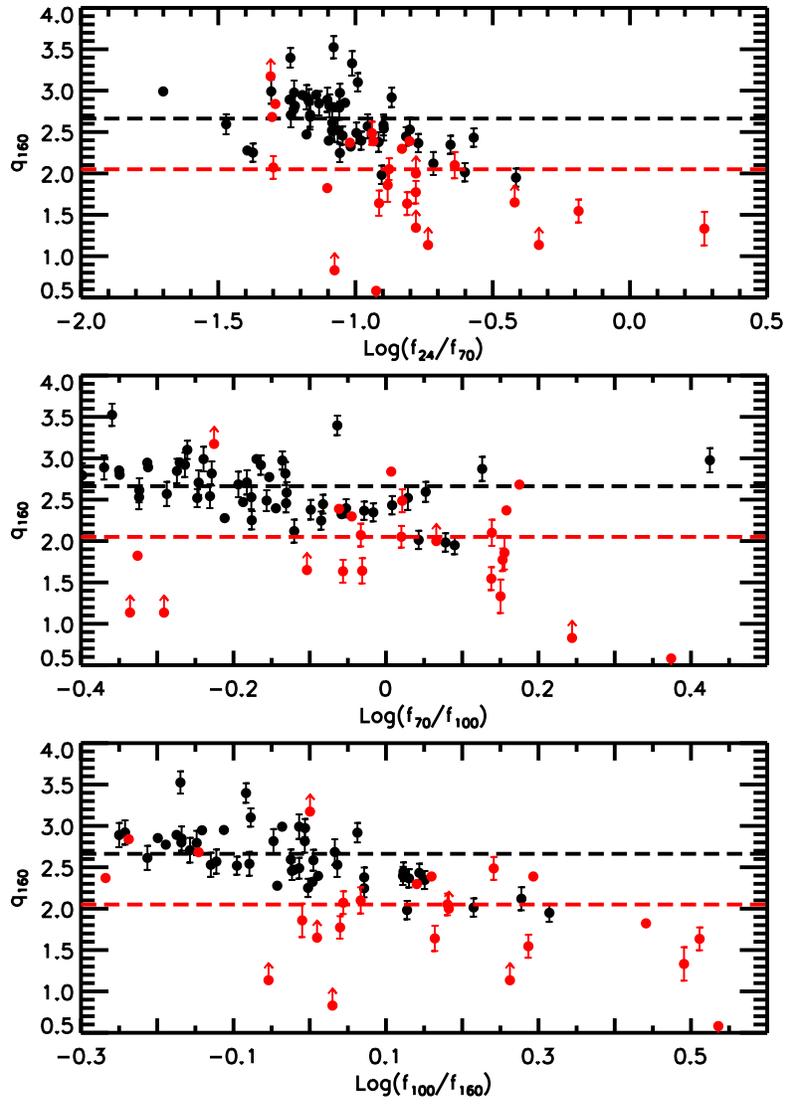}
\caption{The same as Fig.~\ref{Fcolor24} but for  q$_{\rm 160{\mu}m}$.\label{Fcolor160}}
\end{figure}

The  $q_{160{\mu}m}$  as a  function  of  the  IR  color is  shown  in
Figure~\ref{Fcolor160}.  An  inverse relation of  $q_{160{\mu}m}$ with
$f_{24{\mu}m}/f_{70{\mu}m}$ is present with a correlation coefficiency
of    -0.63.     The     relation    between    $q_{160{\mu}m}$    and
$f_{70{\mu}m}/f_{100{\mu}m}$  is  also  apparent, with  a  correlation
coefficiency of  -0.43.  The relationship between  $q_{160{\mu}m}$ and
$f_{100{\mu}m}/f_{160{\mu}m}$ is the strongest  one with a correlation
coefficiency  of -0.71,  indicating  that the  low $q_{160{\mu}m}$  of
metal    poor    galaxies    is    associated    with    their    high
$f_{100{\mu}m}/f_{160{\mu}m}$.

 In  a  summary,  for  the   whole  sample  including  metal-rich  and
 metal-poor galaxies, only $q_{160{\mu}m}$  shows  good relationships
 with three  IR color.  Previous  studies of metal-rich  galaxies also
 found  a  relatively  tight  relationship  between  $q$  at  long  IR
 wavelength  and  the  dust temperature  \citep{Smith14}.   Our  study
 further extends this into metal-poor galaxies.

\section{Discussion}
\label{sect:discussion}

Our multi-wavelength investigations  of $q_{\rm IR}$ at  24 $\mu$m, 70
$\mu$m, 100 $\mu$m, 160 $\mu$m and total far-IR band indicate that the
average  values of  all  $q_{\rm IR}$  except  for $q_{24{\mu}m}$  are
reduced at 12 + log(O/H) $<$ 8.1  as compared to those above 12 + log(O/H)
$=$ 8.1.  At longer IR wavelengths  from 24 $\mu$m to  160 $\mu$m, the
offsets become larger and the trend of decreasing $q_{\rm IR}$ with the IR/FUV ratio and IR
color becomes stronger. In the following, we discusses possible causes
for the observed behavior of $q_{\rm IR}$ at low metallicity.

\begin{figure}
\epsscale{0.70}    
\plotone{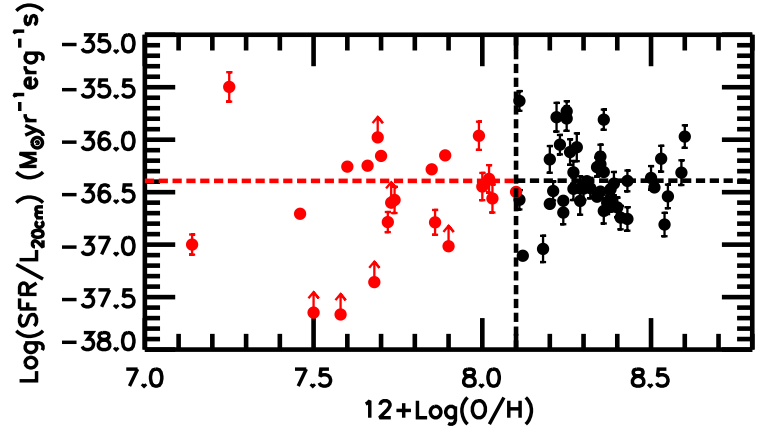}
        \caption{The SFR-to-radio ratio as a function of the oxygen abundance, where
        the SFR includes both obscured and unobscured contributions.\label{F_SFRvsM}}
\end{figure}

As shown  in Fig.~\ref{F_qvsLIR_FUV},  low $q_{\rm IR}$  of metal-poor
galaxies is  associated with low  IR-to-FUV ratio, especially  at long
wavelength (160  $\mu$m) where  a relation  of the  decreasing $q_{\rm
  IR}$ with  decreasing IR-to-FUV ratio  is present.  As  argued below
this   low   $q$   is   driven   by   a   combined   effect   of   low
obscured-SFR/total-SFR  ratio  and  warm   dust  color  in  metal-poor
galaxies.   Although  statistical  studies show  that  low-metallicity
galaxies  have  on  average  similar  dust-to-stellar  mass  ratio  to
metal-rich ones both globally  and locally \citep{Hunt14, Zhou16}, the
$f_{160{\mu}m}$-to-FUV signals a low fraction of obscured SFR relative
to total  SFR.  The low  $f_{160{\mu}m}$-to-FUV ratio results  in that
the portion of  radiation from massive stars that is  absorbed by dust
and re-emitted in  the IR is reduced in metal-poor  galaxies so that a
larger portion  of radiation escape through  far-UV photons.  However,
it is still not enough  to explain the wavelength-dependent offsets of
$q$,  meaning that  at  longer  IR wavelength  the  $q$ of  metal-poor
galaxies  is   lower  than   metal-rich  galaxies.    This  wavelength
dependence needs  to invoke the  efficient dust heating  that increase
the IR emission at short  wavelength, which is consistent with finding
that  the   dust  of   metal-poor  galaxies   is  in   general  warmer
\citep{Remy-Ruyer13,  Zhou16}.  The  efficient  dust  heating is  also
consistent with the observed inverse relationship between $q_{\rm IR}$
and  the IR  color at  long wavelength  ($q_{160{\mu}m}$) as  shown in
Fig.~\ref{Fcolor100}  and  Fig.~\ref{Fcolor160}.  The  above  scenario
implies that the SFR/radio ratio  of metal-poor galaxies should be the
same as that of metal-rich ones,  which is demonstrated by our data as
shown  in  Fig.~\ref{F_SFRvsM}.  The  mean  value  of  log(SFR/L$_{\rm
  20cm}$) were  -36.39$\pm$0.1 and -36.39$\pm$0.05 for  metal-poor and
metal-rich  samples,  respectively.  Here  we  adopted  the  same  SFR
calibration   for   both   samples.   If   any   metallicity-dependent
calibration exits, the result may change.

If the magnetic field is stronger  in metal-poor galaxies, the 1.4 GHz
synchrotron emission  may be boosted  to reduce their $q$  values. The
total  field strength  is measured  by assuming  equipartition between
magnetic  and cosmic-ray  energy.  Such  equipartition magnetic  field
strength of  spiral galaxies is around  10 - 20 $\mu$G  as reviewed by
\citet{Fletcher10}, while  gas-rich galaxies  with high SFRs  can have
much higher strength up to  50 - 100 $\mu$G \citep{Beck05, Adebahr13}.
The magnetic  strength has  been reported for  some of  our metal-poor
sample.  NGC 1569 was measured to have a strength of 14 $\pm$ 3 $\mu$G
\citep{Kepley10}.  The field strength of  blue compact dwarf IZw 18 is
about  11 $\mu$G  or even  higher if  the emitting  region is  compact
\citep{Hunt05}.  SBS 0335-052E has a field strength at least 30 $\mu$G
\citep{Hunt04}.  IC  2574 is  shown to  have a  weak strength  about 4
$\mu$G \citep{Chyzy07}.  Literature studies  show that dwarfs can have
a range  of magnetic field  of strength, $<$ 5  $\mu$G \citep{Chyzy11}
for dwarf  irregulars and 2 -  6 times higher for  blue compact dwarfs
\citep{Hunt04, Hunt05}.  It is  argued that the equipartition magnetic
strength is proportional to the SFR surface density with a power index
around  0.3  for  both  spiral  and  dwarf  galaxies  \citep{Niklas97,
  Chyzy11}. The mechanisms  may be related to the dynamo  as driven by
SFR-related events to amplify the field strength.

\begin{figure}
\epsscale{0.90}    
\plotone{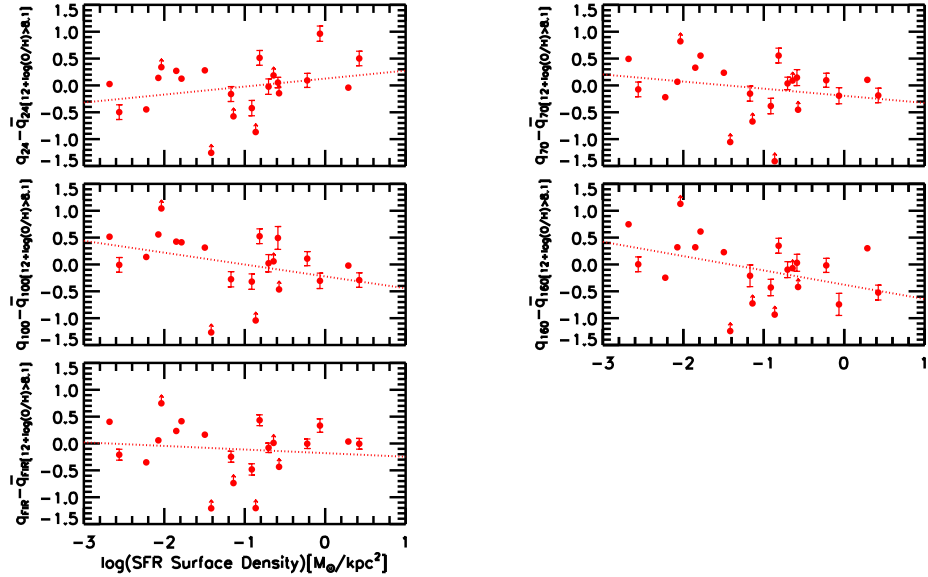}
        \caption{The  deviation  of  $q_{\rm IR}$ 
        of  metal-poor galaxies  from the mean  of metal-rich
        galaxies as a function of the SFR surface densities.
        The red dotted line are the best linear fits.
        \label{f_qvsSFR_density}}
\end{figure}

We test the deviation of the  $q_{\rm IR}$ for our metal-poor galaxies
from the mean value of metal-rich  galaxies as a function of their SFR
surface  densities  as  shown  in  Figure~\ref{f_qvsSFR_density}.   If
magnetic fields  play roles in  lowering q of metal-poor  galaxies, we
should  see  $q_{\rm IR}$  at  all  IR  wavelength decrease  with  the
increasing SFR surface  density, which is nevertheless  not evident in
the  figure.  At  24 $\mu$m,  the offset  of $q_{\rm  IR}$ shows  some
positive relationship with the SFR  surface density with a correlation
coefficiency  of 0.37.   The offset  of $q_{\rm  70{\mu}m}$ is  almost
independent  with   the  SFR  surface  density,  with  a  correlation
coefficiency of -0.16.  Some inverse trends  are seen in the 100 $\mu$m
and 160  $\mu$m panels,  with correlation  coefficiencies of  -0.48 and
-0.46, respectively.   All these  behaviors are  inconsistent  with the
scenario invoking  the increased  magnetic fields which  should reduce
$q_{\rm IR}$ with the same amount  at all IR wavelengths.  Instead the
behavior seen in Figure~\ref{f_qvsSFR_density}  is consistent with the
efficient dust  heating at  high SFR surface  densities to  increase the
short  wavelength   emission  while   reducing  the   long  wavelength
radiation.  As a result, Figure~\ref{f_qvsSFR_density} indeed supports
the  scenario of  low obscured  SFR fraction  and warm  dust color  of
metal-poor  galaxies that  cause  the low  $q_{\rm  IR}$ as  discussed
above.  The  median value of  SFR surface densities of  our metal-poor
galaxies  correspond  to a  magnetic  field  strength about  8  $\mu$G
\citep{Chyzy11}, which is not  larger than the typical  value of
  spiral   galaxies  (10-20   $\mu$G,  \citet{Fletcher10}),   further
supporting it is not the enhanced magnetic field strength that reduces
the $q$ of  metal-poor galaxies.  As compared  to the theoretical
  model \citep{Schleicher16},  our SFR  surface density is  above the
threshold   (10$^{-4}$   -   10$^{-6}$   M$_{\odot}$/yr)   where   the
magnetic-field/SFR  and  cosmic  diffusion  loss may  break  down  the
IR-radio    relationship   of    dwarf    galaxies.    In    addition,
Fig.~\ref{F_SFRvsM} shows that,  if the SFR  calibration does not
  depend  on  the  metallicity,   metal-poor  galaxies  have  similar
SFR-to-radio ratio  to metal-rich ones, also  disfavoring the scenario
of enhanced magnetic field strength.

The  above discussions  assume  1.4 GHz  is  dominated by  non-thermal
synchrotron emission.  Compared to  spiral galaxies, metal-poor dwarfs
have a higher contribution from  thermal emission, but usually no more
than 30\% at  1.4 GHz based on studies of  individual galaxies as well
as statistical studies \citep{Niklas97,  Hunt04}. The low obscured SFR
fraction  means that  more ionized  photons could  escape to  ionize a
larger portion  of gas,  resulting in  enhanced thermal  emission from
free electron. However, if the radio  emission is elevated by 30\% due
to thermal contribution, the $q_{\rm IR}$  is reduced by only 0.1 dex,
not  enough to  explain the  observed  offsets.  And  more over,  such
offsets should be  IR wavelength independent, which  is inconsistent
with what is observed.

As a summary, metal-poor galaxies  have lower $q_{\rm IR}$ as compared
to metal-rich galaxies, with larger offsets in $q_{\rm IR}$ at longer
IR wavelength, which could be explained by the combined effects of low
obscured  SFR  fraction  and   warm  dust  temperature  of  metal-poor
galaxies.   Other  mechanisms such  as  enhanced  magnetic fields  and
enhanced thermal emission  as discussed above, as well  as others such
as more powerful supernovae from more massive stars born in metal-poor
gas  are unable  to  reconcile the  wavelength-dependent behavior  of
$q_{\rm IR}$.

\section{Conclusion}
\label{sect:conclusion}

We have compiled  a sample of 26 metal-poor galaxies  at 12 + log(O/H)
$\leqslant$ 8.1  with multiple IR  and radio 1.4 GHz  photometry data.
With  a comparison  sample  of metal-rich  galaxies  at 12  +
log(O/H) $>$ 8.1,  we perform studies of  IR-radio relationships.  Our
main conclusions are the following:

1.   The  $q_{\rm  IR}$  of  metal-poor galaxies  is  lower  than  the
metal-rich ones,  with larger  offsets at  longer IR  wavelengths. The
mean  offsets  of metal-poor  galaxies  from  metal-rich galaxies  are
-0.06$\pm$0.13,  -0.26$\pm$0.1,   -0.44$\pm$0.12,  -0.61$\pm$0.13  and
-0.24$\pm$0.1  for  $q_{\rm  24{\mu}m}$, $q_{\rm  70{\mu}m}$,  $q_{\rm
  100{\mu}m}$, $q_{\rm 160{\mu}m}$ and $q_{\rm FIR}$, respectively.

2. The  drop in  $q_{\rm IR}$  at long IR  wavelength (160  $\mu$m) is
related  to the  metallicity, the  IR-to-FUV  ratio and  the IR  color
including   $f_{24{\mu}m}/f_{70{\mu}m}$,  $f_{70{\mu}m}/f_{100{\mu}m}$
and $f_{100{\mu}m}/f_{160{\mu}m}$.

3. The total SFR to radio  ratio of metal-poor galaxies have a similar
mean to that of metal-rich ones.

4. The plausible mechanism to explain the behavior of the $q_{\rm IR}$
of metal-poor galaxies invokes the  combined effect of low obscured-SFR/total-SFR
ratio and warm IR color. The former means less absorbed radiation from
massive stars by dust and thus less re-emitted IR emission. The latter
increases  the short  IR wavelength  emission,  in particular  24
  micron, relative  to longer  wavelengths. Other mechanisms  such as
enhanced magnetic field strength and enhanced thermal contribution are
difficult to explain the IR wavelength dependence of the behavior.

\acknowledgements

We  thank   the  anonymous   referee  for  helpful   and  constructive
suggestions that  improved the quality  of the paper.  J.Q.,  Y.S. and
L.Z.  acknowledges  support for  this work  from the  National Natural
Science Foundation  of China (NSFC  grant 11373021) and  the Excellent
Youth   Foundation  of   the  Jiangsu   Scientific  Committee   (grant
BK20150014). This research has made use of the NASA/IPAC Extragalactic
Database (NED)  which is  operated by  the Jet  Propulsion Laboratory,
California Institute  of Technology, under contract  with the National
Aeronautics and Space Administration.






\begin{figure}
\epsscale{0.70}    
\plotone{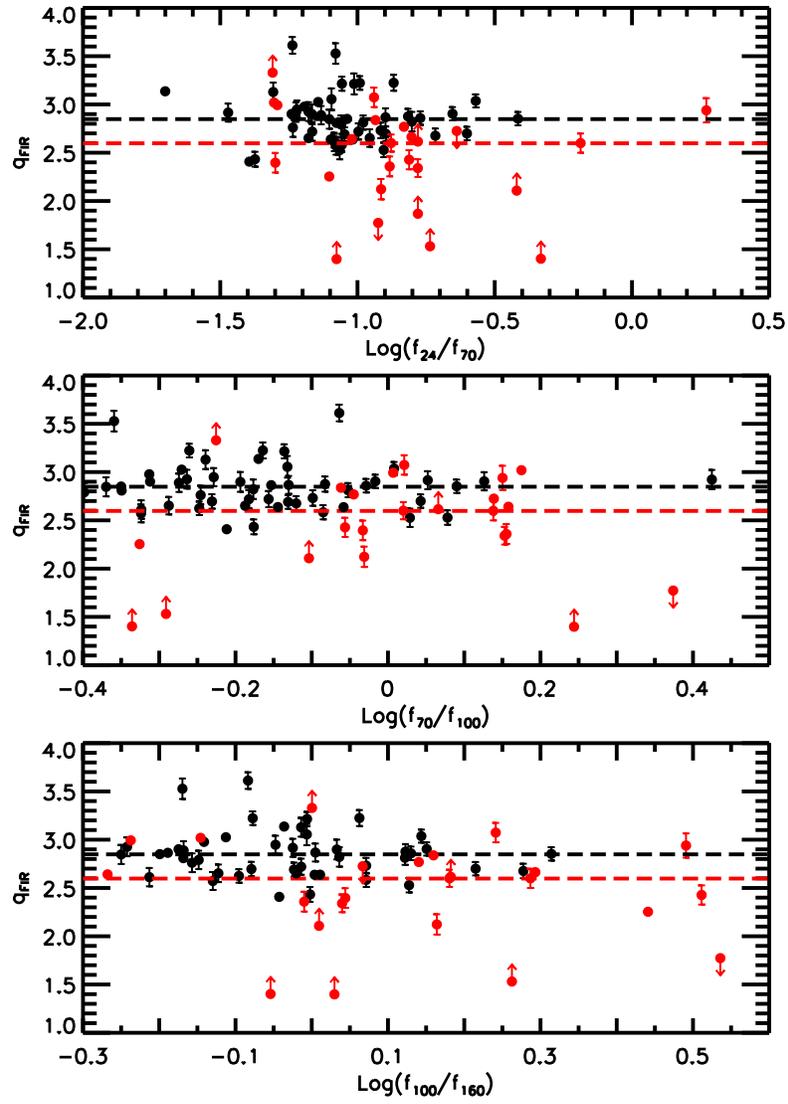}
\caption{The same as Fig.~\ref{Fcolor24} but for  q$_{\rm FIR}$.\label{FcolorFIR}}
\end{figure}

\clearpage


\begin{table}
\tiny
\centering
\caption{The sample of metal-poor galaxies with 12 + log(O/H) $<$ 8.1.}
\label{table_dwarfsample}
\begin{tabular}{lllllllllllll} 
\hline \hline 
Name           & RA[J2000]     & DEC[J2000]       &  D    & FUV$^{a}$      &  REF$^{b}$           & 20cm$^{c}$             & REF$^{d}$ & 12+(O/H)       & REF$^{e}$ &  SFR               &  $\Sigma_{\rm SFR}$    \\
               &               &                  & [Mpc]  & [mJy]  &  & [mJy] & & & & [M$_{\sun}$/yr] & [M$_{\sun}$/yr/kpc$^{2}$] \\
\hline 
SBS0335$-$052E &  03h37m44.06s &  $-$05d02m40.0s  &  56.0      & 0.579$\pm$0.0192    &  1      &           0.46$\pm$0.061  & H04     & 7.25$\pm$0.01    &  1,6      &    8.25$\pm$2.43              &        0.86        \\
NGC1569        &  04h30m49.06s &  $+$64d50m53.0s  &  3.10      & 2.56$\pm$0.024      &  2      &           339$\pm$11      & C98     & 8.02$\pm$0.02    &  1,11     &    2.47$\pm$0.74              &        0.59        \\
Mrk1089        &  05h01m37.76s &  $-$04d15m28.0s  &  56.6      & 5.56$\pm$0.56       &  ...    &           31.7            & H02     & 8.10$\pm$0.08    &  1,13     &    57.9$\pm$16.9              &        1.89        \\
NGC2366        &  07h28m54.66s &  $+$69d12m57.0s  &  3.20      & 28.1$\pm$1.03       &  2      &           19.9            & C02     & 7.70$\pm$0.01    &  1,16     &    0.26$\pm$0.07              &       0.014        \\
IC2233         &  08h13m58.91s &  $+$45d44m31.7s  &  9.48      & 2.81$\pm$0.0258     &  2      &        $<$1               & L05     & 7.69$\pm$0.00    &  39       &    0.17$\pm$0.04              &      0.0086        \\
HolmbergII     &  08h19m04.98s &  $+$70d43m12.1s  &  3.28      & 38.0$\pm$0.35       &  2      &           30.9$\pm$3      & VLA     & 7.72$\pm$0.14    &  5        &    0.10$\pm$0.02              &      0.0028        \\
DDO053         &  08h34m07.20s &  $+$66d10m54.0s  &  3.31      & 2.01$\pm$0.019      &  2      &           1.1             & L05     & 7.60$\pm$0.11    &  5        &    0.012$\pm$0.003            &      0.0084        \\
UGC4483        &  08h37m03.00s &  $+$69d46m31.0s  &  3.20      & 1.82$\pm$0.084      &  3      &           1.1             & H02     & 7.46$\pm$0.02    &  1,19     &    0.004$\pm$0.0007           &       0.006        \\
IZw18          &  09h34m02.03s &  $+$55d14m28.0s  &  18.2      & 1.36$\pm$0.013      &  2      &           1.79$\pm$0.18   & C05     & 7.14$\pm$0.01    &  1,22     &    0.11$\pm$0.02              &         ...        \\
SBS0940+544    &  09h44m16.61s &  $+$54d11m34.3s  &  23.0      & 0.148$\pm$0.015     &  ...    &        $<$2.3             & H02     & 7.50$\pm$0.00    &  17,23    &    0.049$\pm$0.015            &         ...        \\
IC2574         &  10h28m23.48s &  $+$68d24m43.7s  &  3.56      & 3600$\pm$33         &  2      &           10.7            & C02     & 7.85$\pm$0.14    &  5        &    0.13$\pm$0.03              &       0.002        \\
Mrk153         &  10h49m05.03s &  $+$52d20m08.0s  &  40.3      & 2.98$\pm$0.0624     &  1      &           4.50$\pm$0.6    & C98     & 7.86$\pm$0.04    &  1,26     &    2.14$\pm$0.51              &        0.12        \\
VIIZw403       &  11h27m59.90s &  $+$78d59m39.0s  &  4.50      & 2.91$\pm$0.0268     &  2      &           1.2             & T04     & 7.66$\pm$0.01    &  1,27     &    0.025$\pm$0.006            &       0.016        \\
Mrk1450        &  11h38m35.78s &  $+$57d52m27.0s  &  19.8      & ...                 &  ...    &        $<$2               & H02     & 7.84$\pm$0.01    &  1,28     &    0.74$\pm$0.22$^{\star}$     &        0.23        \\
UM461        &  11h51m33.35s &  $-$02d22m22.0s  &  13.2      & 0.525$\pm$0.0484    &  4      &        $<$2.6             & H02     & 7.73$\pm$0.01    &  1,26     &    0.20$\pm$0.06              &        0.27        \\
UM462        &  11h52m37.19s &  $-$02d28m09.9s  &  13.4      & 2.77$\pm$0.255      &  4      &           6.4$\pm$0.6     & C98     & 8.00$\pm$0.00    &  17,35    &    0.74$\pm$0.21              &        0.21        \\
SBS1159+545    &  12h02m02.47s &  $+$54d15m50.0s  &  57.0      & ...                 &  ...    &        $<$2.3             & H02     & 7.44$\pm$0.01    &  1,28     &    0.62$\pm$0.19$^{\star}$     &        0.14        \\
SBS1211+540    &  12h14m02.58s &  $+$53d45m17.0s  &  19.3      & 0.169$\pm$0.0098    &  1      &        $<$2.9             & H02     & 7.58$\pm$0.01    &  1,28     &    0.04$\pm$0.01              &        1.86        \\
Mrk209         &  12h26m15.92s &  $+$48d29m37.0s  &  5.80      & 3.08$\pm$0.142      &  3      &           4.5$\pm$0.5     & C98     & 7.74$\pm$0.01    &  1,27     &    0.07$\pm$0.02              &       0.068        \\
SBS1249+493    &  12h51m52.50s &  $+$49d03m28.0s  &  111       & 0.101$\pm$0.01      &  ...    &        $<$2.2             & H02     & 7.68$\pm$0.02    &  1,32     &    2.14$\pm$0.61              &         ...        \\
SHOC391        &  12h53m05.97s &  $-$03d12m58.9s  &  99.7      & 1.00$\pm$0.034      &  1      &           6$\pm$0.5       & C98     & 7.99$\pm$0.02    &  40       &    117$\pm$35                 &        2.22        \\
NGC4861        &  12h59m02.34s &  $+$34d51m34.0s  &  7.50      & 12.9$\pm$0.12       &  2      &           10              & T04     & 7.89$\pm$0.01    &  1,27     &    0.72$\pm$0.20              &       0.032        \\
NGC5408        &  14h03m20.91s &  $-$41d22m39.7s  &  4.87      & ...                 &  ...    &         6.53$\pm$0.661    & VLA     & 7.81$\pm$0.09    &  5        &    0.31$\pm$0.09$^{\star}$     &        0.15        \\
Mrk475         &  14h39m05.46s &  $+$36d48m21.9s  &  11.2      & 0.334$\pm$0.014     &  1      &        $<$2.7             & H02     & 7.90$\pm$0.00    &  17,35    &    0.059$\pm$0.017            &       0.072        \\
SBS1533+574    &  15h34m13.80s &  $+$57d17m06.0s  &  54.2      & 0.7$\pm$0.07        &  ...    &           1.4$\pm$0.159   & FIRST   & 8.05$\pm$0.01    &  1,27     &    $<$ 6.28                   &         ...        \\
Mrk1499        &  16h35m21.00s &  $+$52d12m52.3s  &  39.0      & ...                 &  ...    &           1.49$\pm$0.142  & FIRST   & 8.10$\pm$0.00    &  17,37    &    1.57$\pm$0.31$^{\star}$     &        0.26        \\
Mrk930         &  23h31m58.39s &  $+$28h56m49.9s  &  77.8      & 0.589$\pm$0.0543    &  4      &           13.1$\pm$1      & C98     & 8.03$\pm$0.01    &  1,6      &    39.2$\pm$11.7              &         ...        \\
    \hline            
    \hline
\multicolumn{12}{l}{$^{a}$ Far-ultraviolate flux.}\\
\multicolumn{12}{l}{$^{b}$ (1) Galaxy Evolution Explorer (GALEX) All-Sky Catalog based on GALEX General Release 6; (2) \cite{Gil07}; (3) \cite{Lee11}; (4) \cite{Brown14};}\\
\multicolumn{12}{l}{(5) \cite{Munoz09}; (6) \cite{Hao11}}\\
\multicolumn{12}{l}{$^{c}$ 20 cm radio continuum emission flux.}\\
\multicolumn{12}{l}{$^{d}$ The reference of 20cm flux:  C02 \citep{Condon02}; C98  \citep{Condon98}; K91 \citep{Klein91}; H02 \citep{Hopkins02}; H94 \citep{Hunter94}; }\\
\multicolumn{12}{l}{H04 \citep{Hunt04}; C04 \citep{Cannon04}; M08 \citep{Matthews08}; L05 \citep{Leroy05}; C05 \citep{Cannon05}; T04 \citep{Thuan04}; }\\
\multicolumn{12}{l}{Kepley11 \citep{Kepley11}; H64 \citep{Heeschen64}; S76 \citep{Sulentic76}; VLA (http://archive.nrao.edu/nvas/); FIRST (http://sundog.stsci.edu/index.html).}\\
\multicolumn{12}{l}{$^{e}$ the reference of oxygen abundace:
(1) \citep{Remy-Ruyer13}; (2) \citep{Magrini09}; (3) \citep{Ugryumov03}; (4) \citep{Guseva12};}\\
\multicolumn{12}{l}{(5) \citep{Moustakas10}; (6) \citep{Izotov98}; (7) \citep{Skillman03}; (8) \citep{vanZee96}; (9) \citep{Kobulnicky99b}; (10) \citep{Izotov04};}\\
\multicolumn{12}{l}{ (11) \citep{Kobulnicky97}; (12) \citep{Lee04}; (13) \citep{Lopez-Sanchez04}; (14) \citep{Guseva00}; (15) \citep{Masegosa94}; }\\
\multicolumn{12}{l}{(16) \citep{Saviane08}; (17) \citep{Wu08}; (18) \citep{Pustilnik03}; (19) \citep{vanZee06}; (20) \citep{Heckman98}; }\\
\multicolumn{12}{l}{(21) \citep{Kobulnicky99a}; (22) \citep{Izotov99a}; (23) \citep{Thuan05}; (24) \citep{Storchi-Bergmann94}; (25) \citep{Kong02}; }\\
\multicolumn{12}{l}{(26) \citep{Izotov06}; (27) \citep{Izotov97}; (28) \citep{Izotov94}; (29) \citep{Izotov07}; (30) \citep{McCall85}; (31) \citep{Popescu00};}\\
\multicolumn{12}{l}{(32) \citep{Thuan95}; (33) \citep{Guseva07}; (34) \citep{Guseva03b}; (35) \citep{Izotov99b}; (36) \citep{Guseva03a}; (37) \citep{Shi05};}\\
\multicolumn{12}{l}{(38) \citep{Lee06}; (39) \citep{Berg12}; (40) \citep{Esteban14}.}\\
\multicolumn{12}{l}{$^{\star}$ SFR estimate without FUV (no FUV archival data).}\\
\hline 
\end{tabular}
\end{table}

\clearpage


\begin{table}
\scriptsize
\centering
\caption{The IR photometry of metal-poor galaxies.}
\label{table_dwarfinfrared}
\begin{tabular}{lcccc} 
\hline \hline 
Name         &  24$\mu$m   & 70$\mu$m[mJy]   & 100$\mu$m[mJy]    & 160$\mu$m[mJy]     \\
             &  [mJy]      & [mJy]         &    [mJy]      &  [mJy]           \\                       
\hline
SBS0335-052E &   81$\pm$8       &  42$\pm$5       &    30$\pm$3       &   10$\pm$2         \\
NGC1569      &   8038$\pm$804   &  60628$\pm$6063 &    57877$\pm$5778 &   38187$\pm$3820   \\
Mrk1089	     &   551$\pm$55     &  5776$\pm$580   &    4015$\pm$402   &   7438$\pm$786     \\
NGC2366      &   710$\pm$71     &  6106$\pm$611   &    7038$\pm$704   &   4873$\pm$491     \\
IC2233       &   44$\pm$4       &  887$\pm$89     &    1491$\pm$149   &   1490$\pm$153     \\
HolmbergII   &   188$\pm$19     &  3741$\pm$374   &    4036$\pm$404   &   3647$\pm$370     \\
DDO053	     &   29$\pm$3       &  185$\pm$20     &    529$\pm$54     &   269$\pm$40       \\
UGC4483	     &   7.5$\pm$0.8    &  95$\pm$10      &    201$\pm$21     &   73$\pm$15        \\
IZw18	     &   6.6$\pm$0.7    &  55$\pm$6       &    23$\pm$2       & $<$6.8             \\
SBS0940+544  &   2.5$\pm$0.4    &  ...            &    ...            &   ...              \\
IC2574	     &   222$\pm$22     &  4343$\pm$436   &    4275$\pm$429   &   7384$\pm$752     \\
Mrk153	     &   33$\pm$3       &  267$\pm$27     &    287$\pm$30     &   197$\pm$24       \\
VIIZw403     &   31$\pm$3       &  617$\pm$65     &    412$\pm$44     &   577$\pm$124      \\
Mrk1450      &   59$\pm$6       &  354$\pm$36     &    304$\pm$31     &   200$\pm$23       \\
UM461	     &   35$\pm$4       &  93$\pm$9       &    119$\pm$13     &   116$\pm$15       \\
UM462	     &   121$\pm$12     &  787$\pm$79     &   896$\pm$143$^I$ &   276$\pm$28       \\
SBS1159+545  &   6$\pm$0.6      &  13$\pm$3       &    28$\pm$4       &   31$\pm$7         \\
SBS1211+540  &   3$\pm$0.3      &  37$\pm$4       &    21$\pm$3       &   20$\pm$5         \\
Mrk209	     &   59$\pm$6       &  454$\pm$47     &    317$\pm$34     &   325$\pm$82       \\
SBS1249+493  &   5$\pm$0.5      &  28$\pm$4       &    55$\pm$7       &   30$\pm$8         \\
SHOC391      &   364$\pm$36     & 560$\pm$56      &    408$\pm$41     &   211$\pm$23       \\
NGC4861	     &   365$\pm$36     &  2471$\pm$247   &    2740$\pm$274   &   1983$\pm$200     \\
NGC5408      &   421$\pm$42     &  3664$\pm$367   &    3489$\pm$349   &   2002$\pm$204     \\
Mrk475	     &   14$\pm$1       &  83$\pm$12      &    ...            &   60$\pm$14        \\
SBS1533+574  &   $<$65          &  283$\pm$29     &    205$\pm$22     &   176$\pm$26       \\
Mrk1499	     &   32.2$^I$       &  402$\pm$40     &   617$\pm$173$^I$ &   ...              \\
Mrk930       &   201$\pm$20     &  1211$\pm$121   &    851$\pm$86     &   777$\pm$80       \\
\hline\hline
\multicolumn{5}{l}{$^I$ Taken from IRAS.}\\
\hline\\
\end{tabular}
\end{table}
\clearpage


\begin{table}
\fontsize{5}{5}\selectfont
\caption{The comparison sample of metal-rich galaxies with 12 + log(O/H) $>$ 8.1.}
\label{table_comparionsinfrared} 
\begin{tabular}{llllllllllllll}
\hline \hline  \\[-3ex]
Name          & D      & 3.6$\mu$m          & 24$\mu$m          & 70$\mu$m            & 100$\mu$m               & 160$\mu$m           & FUV               & REF  & 20cm          & REF    & 12+log(O/H)          & REF     \\ 
              & [Mpc]  & [Jy]  & [Jy] & [Jy] & [Jy] & [Jy] &  [mJy] &   & [mJy]  &  & \\
\hline \\[-3ex]                                                                                                                                                                             
IC10          &  0.7     &      ...                &    2.79$\pm$0.17        &     140$\pm$7           &        207$\pm$10           &      225$\pm$11         &     ...                & ...  &      230            & C02    &    8.17$\pm$0.03     & 1,2     \\
Haro11        &  92.1    &      22.6$\pm$0.678     &    2.36$\pm$0.0473      &    6.14$\pm$0.31        &       4.99$\pm$0.25         &     2.42$\pm$0.12       &     2.74$\pm$0.0225    & 1    &     27.2$\pm$0.9    & C98    &    8.36$\pm$0.01     & 1,4     \\
NGC0337       &  21      &      98.6$\pm$0.227     &    0.55$\pm$0.0506      &      13$\pm$0.7         &       19.5$\pm$1            &     19.6$\pm$1          &     4.48$\pm$0.0205    & 1    &      110$\pm$4.1    & C98    &    8.18$\pm$0.07     & 5     \\
UM311         &  23.5    &      ...                &    ...                  &    2.94$\pm$0.15        &       5.63$\pm$0.28         &      6.1$\pm$0.31       &     ...                & ...  &      ...            & ...    &    8.36$\pm$0.01     & 1,6     \\
NGC625        &  3.9     &      123 $\pm$17        &   0.879$\pm$0.095       &    6.49$\pm$0.32        &       9.47$\pm$0.47         &      8.2$\pm$0.41       &     11.8$\pm$0.544     & 3    &      9.9$\pm$1      & C04    &    8.22$\pm$0.02     & 1,7     \\
NGC0855       &  9.73    &      42.6$\pm$0.294     &  0.0777$\pm$0.00715     &    2.30$\pm$0.12        &       2.04$\pm$0.12         &     2.16$\pm$0.12       &     1.08$\pm$0.01      & 2    &      5.5$\pm$0.6    & C98    &    8.29$\pm$0.10     & 5      \\
NGC0925       &  8.58    &      321 $\pm$1.48      &   0.899$\pm$0.0145      &    10.8$\pm$0.6         &       24.7$\pm$1.2          &     36.5$\pm$1.8        &     29.4$\pm$0.271     & 2    &     10.9$\pm$2      & C98    &    8.25$\pm$0.01     & 5      \\
NGC1097       &  17.1    &      1240$\pm$170       &    6.63$\pm$0.27        &   59.84$\pm$4.66        &        116$\pm$6            &    153.8$\pm$18.5       &     29.9$\pm$1.1       & 2    &      415$\pm$42     & D07    &    8.55$\pm$0.09     & 5     \\
NGC1140       &  20      &      ...                &   0.388$\pm$0.00782     &    4.04$\pm$0.2         &       4.62$\pm$0.23         &     4.58$\pm$0.23       &     8.87$\pm$0.0817    & 2    &     21.7            & H94    &    8.38$\pm$0.01     & 1,10     \\
NGC1482       &  19.6    &      201 $\pm$1.86      &    3.57$\pm$0.0329      &    40.7$\pm$2           &       49.5$\pm$2.5          &       42$\pm$2.1        &    0.302$\pm$0.0222    & 2    &     238 $\pm$8.4    & C98    &    8.11$\pm$0.13     & 5      \\
NGC1705       &  5.1     &      27.3$\pm$0.126     &   0.056$\pm$0.002       &    1.37$\pm$0.07        &       1.46$\pm$0.07         &      1.1$\pm$0.06       &     13.6$\pm$0.0886    & 1    &      ...            & ...    &    8.27$\pm$0.11     & 1,12     \\
IIZw40        &  12.1    &        16$\pm$0.481     &     1.6$\pm$0.032       &    6.39$\pm$0.32        &       5.79$\pm$0.29         &     3.53$\pm$0.18       &     ...                & ...  &     34.2$\pm$1.4    & C98    &    8.23$\pm$0.01     & 1,14     \\
NGC2403       &  3.13    &      1880$\pm$250       &    5.84$\pm$0.24        &   86.36$\pm$6.18        &      64.59$\pm$3.23         &    245.6$\pm$29.6       &      192$\pm$1.77      & 2    &      330$\pm$33     & D07    &     8.3$\pm$0.14     & 5     \\
UGC4274       &  6.9     &      80.8$\pm$2.43      &   0.276$\pm$0.0254      &    4.04$\pm$0.203       &       6.31$\pm$0.2          &     5.85$\pm$0.703      &     6.79$\pm$0.0626    & 2    &     12.1$\pm$1.9    & M08    &    8.5               & 17     \\
He2-10        &  8.7     &      97.4$\pm$2.92      &    5.68$\pm$0.114       &    25.6$\pm$1.3         &       26.6$\pm$1.3          &     18.8$\pm$0.9        &     ...                & ...  &     84.7$\pm$3.4    & C98    &    8.43$\pm$0.01     & 1,9     \\
NGC2798       &  26.4    &      69.2$\pm$0.319     &    2.54$\pm$0.0176      &    24.2$\pm$1.2         &       27.3$\pm$1.4          &     20.6$\pm$1          &    0.964$\pm$0.00888   & 2    &     82.8$\pm$3      & C98    &    8.34$\pm$0.08     & 5     \\
NGC2903       &  8.9     &      1130$\pm$33.9      &    9.69$\pm$0.194       &    76.4$\pm$3.83        &        130$\pm$0.2          &      156$\pm$20.3       &     41.3$\pm$0.38      & 2    &      448$\pm$14     & C98    &    9.3               & 17,21     \\
NGC2976       &  3.95    &       408$\pm$0.94      &    1.38$\pm$0.00317     &    19.2$\pm$1           &       35.8$\pm$1.8          &     46.4$\pm$2.3        &     21.5$\pm$1.58      & 2    &       52            & C02    &    8.36$\pm$0.06     & 5      \\
NGC3049       &  24.8    &      41.1$\pm$0.0947    &    0.43$\pm$0.02        &    3.40$\pm$0.18        &       4.59$\pm$0.23         &     4.54$\pm$0.24       &     2.3$\pm$0.00373    & 1    &     11.8$\pm$1.7    & C98    &    8.53$\pm$0.01     & 5     \\
NGC3031       &  3.55    &     10920$\pm$1480      &    5.09$\pm$0.2         &   85.18$\pm$5.96        &      32.03$\pm$1.28         &      360$\pm$43.4       &      100$\pm$0.921     & 2    &      380$\pm$38     & D07    &    8.37$\pm$0.14     & 5     \\
HoIX          &  3.7     &         7$\pm$1         &   0.004$\pm$0.0006      &   0.054$\pm$0.014       &      ...                    &    0.204$\pm$0.041      &     2.27$\pm$0.0209    & 2    &      ...            & ...    &    8.14$\pm$0.11     & 5     \\
NGC3077       &  3.8     &       496$\pm$14.9      &    1.79$\pm$0.0362      &    20.4$\pm$1           &       27.9$\pm$1.4          &     28.3$\pm$1.4        &     ...                & ...  &     30.1$\pm$1.5    & C98    &    8.6               & 17,24     \\
NGC3184       &  12.2    &       515$\pm$1.19      &    1.42$\pm$0.00327     &    15.5$\pm$0.8         &       34.7$\pm$1.7          &     54.9$\pm$2.8        &     37.1$\pm$0.0684    & 5    &       77            & C02    &    8.51$\pm$0.01     & 5     \\
NGC3198       &  14      &       270$\pm$0.621     &   0.561$\pm$0.0517      &    9.75$\pm$0.51        &         20$\pm$1            &     29.9$\pm$1.5        &     21.7$\pm$0.2       & 2    &     38.4            & C02    &    8.34$\pm$0.02     & 5     \\
NGC3265       &  1.6     &      25.4$\pm$0.117     &     0.3$\pm$0.01        &    2.47$\pm$0.13        &        3.1$\pm$0.16         &     2.63$\pm$0.15       &    0.474$\pm$0.00437   & 2    &       11$\pm$1      & C98    &    8.27$\pm$0.14     & 5     \\
Haro2         &  21.7    &      26.1$\pm$0.18      &   0.845$\pm$0.00973     &    4.99$\pm$0.25        &       5.33$\pm$0.27         &     3.95$\pm$0.2        &      3.8$\pm$0.035     & 2    &       17$\pm$0.7    & C98    &    8.23$\pm$0.03     & 1,25     \\
NGC3351       &  10.1    &       773$\pm$1.78      &    2.58$\pm$0.12        &    25.3$\pm$1.3         &       46.1$\pm$2.3          &     55.1$\pm$2.8        &     14.5$\pm$0.133     & 2    &     43.6$\pm$2      & C98    &    8.60$\pm$0.01     & 5     \\
Haro3         &  19.3    &      ...                &    0.81$\pm$0.081       &    5.30$\pm$0.26        &       6.41$\pm$0.32         &     4.83$\pm$0.24       &     4.57$\pm$0.0421    & 2    &     17.3$\pm$1.3    & C98    &    8.28$\pm$0.01     & 1,10     \\
NGC3521       &  10.1    &      2050$\pm$280       &    5.51$\pm$0.22        &   63.13$\pm$4.54        &        158$\pm$8            &    222.3$\pm$26.8       &     14.6$\pm$0.134     & 2    &      357$\pm$36     & D07    &    8.38$\pm$0.11     & 5     \\
NGC3621       &  6.55    &       990$\pm$130       &     3.7$\pm$0.19        &   50.21$\pm$3.94        &       94.4$\pm$4.7          &      139$\pm$17.1       &     42.9$\pm$1.97      & 2    &      198$\pm$20     & D07    &    8.29$\pm$0.14     & 5     \\
NGC3773       &  17      &      22.2$\pm$0.307     &    0.13$\pm$0.003       &    1.29$\pm$0.08        &       1.85$\pm$0.11         &     1.91$\pm$0.14       &     4.22$\pm$0.0311    & 5    &      6.2$\pm$0.6    & C98    &    8.43$\pm$0.03     & 5     \\
UM448         &  87.8    &      14.5$\pm$0.434     &   0.644$\pm$0.0129      &    5.17$\pm$0.26        &       4.32$\pm$0.389        &     3.22$\pm$0.17       &     2.03$\pm$0.0209    & 1    &     33.5$\pm$1.4    & C98    &    8.32$\pm$0.01     & 1,6     \\
NGC4194       &  41.5    &      94.8$\pm$2.84      &    3.67$\pm$0.338       &    19.1$\pm$0.953       &       25.2$\pm$0.11         &     13.3$\pm$1.59       &     2.18$\pm$0.2       & 4    &      101$\pm$3.1    & C98    &    8.2               & 17,24     \\
NGC4214       &  2.9     &       312$\pm$9.37      &    1.97$\pm$0.0395      &    24.5$\pm$1.2         &       33.2$\pm$1.6          &     33.7$\pm$1.7        &     91.2$\pm$4.2       & 3    &     51.5$\pm$10.3   & K11    &    8.26$\pm$0.01     & 1,9     \\
NGC4254       &  16.5    &       700$\pm$100       &     4.2$\pm$0.17        &   50.29$\pm$3.6         &        106$\pm$5            &    142.9$\pm$17.2       &     31.1$\pm$1.46      & 6    &      422$\pm$42     & D07    &    8.41$\pm$0.14     & 5     \\
NGC4321       &  14.32   &       950$\pm$130       &    3.34$\pm$0.13        &   40.59$\pm$2.9         &       85.5$\pm$4.3          &    139.6$\pm$16.8       &     3.32$\pm$0.0266    & 1    &      340$\pm$34     & D07    &    8.43$\pm$0.08     & 5     \\
NGC4449       &  4.2     &       493$\pm$14.8      &    3.27$\pm$0.0655      &    49.3$\pm$2.5         &       75.9$\pm$3.8          &     79.5$\pm$4          &      164$\pm$7.57      & 3    &      269            & C02    &     8.2$\pm$0.11     & 1,30     \\
NGC4536       &  15.3    &       418$\pm$2.89      &    3.49$\pm$0.0161      &    38.9$\pm$2           &       52.6$\pm$2.6          &     55.5$\pm$2.8        &     14.9$\pm$0.137     & 2    &      194$\pm$7.6    & C98    &    8.21$\pm$0.08     & 5     \\
NGC4559       &  10.3    &       350$\pm$50        &    1.12$\pm$0.05        &   16.89$\pm$1.2         &         31$\pm$1.6          &    54.15$\pm$6.53       &     47.4$\pm$0.437     & 2    &       65$\pm$7      & D07    &    8.27$\pm$0.10     & 5     \\
NGC4625       &  8.2     &      49.7$\pm$0.8       &   0.114$\pm$0.0105      &    1.36$\pm$0.12        &       3.04$\pm$0.2          &     4.48$\pm$0.23       &      5.3$\pm$0.0488    & 2    &      7.1            & C02    &    8.35$\pm$0.17     & 5     \\
NGC4631       &  5.83    &      1190$\pm$2.75      &    5.53$\pm$0.509       &     137$\pm$7           &        223$\pm$11           &      246$\pm$12         &     71.8$\pm$0.661     & 2    &     1300            & C98    &    8.12$\pm$0.11     & 5     \\
NGC4725       &  11.91   &      1140$\pm$150       &    0.86$\pm$0.04        &    8.85$\pm$0.66        &       22.8$\pm$1.2          &    59.91$\pm$7.36       &     23.1$\pm$0.107     & 5    &       28$\pm$3      & D07    &    8.35$\pm$0.13     & 5     \\
NGC4736       &  5.2     &      3600$\pm$490       &    5.65$\pm$0.23        &   93.93$\pm$7.34        &        159$\pm$8            &    177.4$\pm$21.4       &     59.2$\pm$0.545     & 2    &      271$\pm$27     & D07    &    8.39$\pm$0.08     & 5     \\
NGC4826       &  7.48    &      2520$\pm$340       &    2.72$\pm$0.15        &   55.16$\pm$5.05        &       95.7$\pm$4.8          &    98.82$\pm$12.67      &     10.8$\pm$0.0992    & 2    &      101$\pm$10     & D07    &    8.59$\pm$0.11     & 5     \\
NGC5033       &  14.8    &       640$\pm$90        &    1.97$\pm$0.08        &   28.81$\pm$2.09        &      43.85$\pm$2.63         &    91.07$\pm$11.2       &     18.4$\pm$0.017     & 5    &      178$\pm$18     & D07    &    8.24$\pm$0.24     & 5     \\
NGC5055       &  7.8     &      2380$\pm$320       &    5.73$\pm$0.23        &   72.57$\pm$5.16        &        170$\pm$8            &    302.3$\pm$36.6       &     34.7$\pm$0.319     & 5    &      390$\pm$39     & D07    &    8.38$\pm$0.18     & 5     \\
NGC5194       &  7.62    &      2660$\pm$360       &   12.67$\pm$0.53        &   147.1$\pm$10.6        &      137.7$\pm$8.26         &    494.7$\pm$59.8       &      124$\pm$1.14      & 2    &     1490$\pm$150    & D07    &    8.54$\pm$0.09     & 5     \\
NGC5253       &  4       &       255$\pm$7.64      &    8.87$\pm$0.177       &    32.9$\pm$1.6         &       32.3$\pm$1.6          &     23.2$\pm$1.2        &     31.9$\pm$0.294     & 2    &     85.7$\pm$3.4    & C98    &    8.25$\pm$0.02     & 1,9     \\
NGC5474       &  5.95    &       104$\pm$0.241     &   0.193$\pm$0.00356     &    3.24$\pm$0.18        &       4.61$\pm$0.25         &     7.12$\pm$0.37       &     22.5$\pm$0.207     & 2    &       12            & C02    &    8.31$\pm$0.22     & 5     \\
NGC5713       &  22      &       200$\pm$0.461     &    2.31$\pm$0.0107      &    28.9$\pm$1.4         &       40.3$\pm$2            &     39.3$\pm$2          &     3.87$\pm$0.0357    & 2    &      158            & C02    &    8.24$\pm$0.06     & 5     \\
NGC6822       &  0.5     &      3080$\pm$92.3      &    3.18$\pm$0.13        &    54.9$\pm$2.8         &       63.6$\pm$3.2          &     77.1$\pm$3.9        &      344$\pm$0.316     & 5    &     30.9$\pm$3.26   & VLA    &    8.11$\pm$0.01     & 1,38     \\
NGC6946       &  5.66    &      3290$\pm$7.57      &    20.2$\pm$0.357       &     246$\pm$12          &        435$\pm$22           &      542$\pm$27         &      246$\pm$0.678     & 5    &     1640$\pm$51     & S76    &    8.40$\pm$0.03     & 5     \\
NGC7331       &  14.52   &      1610$\pm$220       &    4.36$\pm$0.25        &   74.97$\pm$6.62        &        132$\pm$7            &    189.5$\pm$24.3       &     8.02$\pm$0.0738    & 2    &      373$\pm$37     & D07    &    8.36$\pm$0.07     & 5     \\
NGC7552       &  21      &       450$\pm$60        &   10.66$\pm$0.44        &   67.59$\pm$11.1        &      101.5$\pm$6.09         &    93.39$\pm$11.25      &     6.98$\pm$0.0643    & 2    &      276$\pm$28     & D07    &    8.35$\pm$0.03     & 5     \\
HS2352+2733   &  117     &      ...                &    ...                  &   0.039$\pm$0.003       &      0.016$\pm$0.002        &    0.016                &   0.0322$\pm$0.00476   & 1    &      ...            & ...    &     8.40$\pm$0.10    & 1,3     \\
NGC7793       &  3.93    &       746$\pm$1.72      &    2.05$\pm$0.00471     &      32$\pm$1.6         &       65.8$\pm$3.3          &     91.1$\pm$4.6        &      108$\pm$0.991     & 2    &      103            & C96    &    8.31$\pm$0.02     & 5     \\ 
\hline\hline 
\multicolumn{12}{l}{1, The IR photometry were retrieved from NED as observed by IRAS, Spitzer and Herschel with the references including}\\
\multicolumn{12}{l}{\citet{Dale07,Dale09,Dale12,Engelbracht08,Munoz09}.}\\
\multicolumn{12}{l}{2, The references for the radio and metallicity data can be found in the Table \ref{table_dwarfsample} notes.}\\
\hline\\
\end{tabular}
\end{table}

\begin{table}
\tiny
\centering
\caption{The mean value of q$_{\rm IR}$ and linear fits of various trends.}
\label{table_linefit}
\begin{tabular}{lccccccc} 
\hline \hline 
name                                            & sample           &                    & 24 $\mu$m     & 70 $\mu$m        &  100 $\mu$m    & 160 $\mu$m    &   FIR      \\
\hline 
                                                &  metal-poor     &     mean     &  1.28$\pm$0.110                &  2.14$\pm$0.084                  & 2.12$\pm$0.109                  & 2.05$\pm$0.122                 & 2.60$\pm$0.077      \\
q                                                &  metal-rich      &     mean      &  1.34$\pm$0.046               &  2.40$\pm$0.036                  & 2.56$\pm$0.042                  & 2.66$\pm$0.049                 & 2.84$\pm$0.036      \\
                                                 &                        &     offset      &  0.06$\pm$0.119               &  0.26$\pm$0.091                  & 0.44$\pm$0.117                  & 0.61$\pm$0.131                 & 0.24$\pm$0.085      \\
\hline
q vs 12+log(O/H)                                &  entire          &  correlation  &  0.164             &  0.318                & 0.417               & 0.567               & 0.347      \\
q vs L$_{\rm IR}$                                &  entire               &  correlation  &  0.127             &  -0.094               & 0.009              & 0.262               & -0.060     \\
q vs f$_{\rm IR}$/f$_{\rm FUV}$                   &   entire                        &  correlation   &  0.224             &  0.052                & 0.249                & 0.420               & 0.037      \\
q vs f$_{24}$/f$_{70}$                           &  entire           &  correlation   &  0.328             &  -0.336               & -0.417               & -0.625             & -0.315     \\
q vs f$_{70}$/f$_{100}$                          &  entire       &  correlation   &  0.200             &  -0.048               & -0.529               & -0.431             & -0.109     \\
q vs f$_{100}$/f$_{160}$                         &  entire       &  correlation    &  0.263             &  -0.208               & -0.332               & -0.710             & -0.294     \\
q-$\overline{\rm q}$ vs SFR surface density     &  metal-poor  &  correlation    &  0.368             &  -0.182               & -0.479               & -0.457             & -0.018     \\        
\hline
                                                                          &  metal-poor     &     mean     &    \multicolumn{2}{c}{-36.394$\pm$0.098} & & &       \\
SFR/L$_{\rm 20cm}$                                        &  metal-rich      &     mean      &    \multicolumn{2}{c}{-36.390$\pm$0.047} & & &       \\
                                                                          &                        &     offset      &  \multicolumn{2}{c}{0.004$\pm$0.108}     & & &        \\
    \hline            
    \hline
\end{tabular}
\tablecomments{Throughout the text, we justified the existence of a weak correlation if the coefficiency is between
0.2 and 0.5, and the existence of a good relationship for the coefficiency larger than 0.5.}
\end{table}

\end{document}